\documentclass[12pt]{article}
\pdfoutput=1
\usepackage{jheppub}
\usepackage{cancel}
\usepackage{amsmath}
\usepackage{amsfonts}
\usepackage{amssymb}
\usepackage{graphicx}

\usepackage{subfiles}
\usepackage{jheppub}
\usepackage{amsmath}
\usepackage{amsfonts}
\usepackage{amssymb}
\usepackage{graphicx}
\usepackage{bm}
\usepackage[capitalise]{cleveref}
\usepackage[export]{adjustbox}
\usepackage{hyperref}
\usepackage{subfiles}
\usepackage{caption}
\usepackage{subcaption}
\usepackage{placeins}
\usepackage{tikz}
\usetikzlibrary{positioning,trees,decorations.pathmorphing,decorations.markings,decorations.pathreplacing,calc,shapes,patterns,arrows}
\usepackage{xcolor}
\definecolor{bluetto}{HTML}{0088ff}
\definecolor{verdino}{HTML}{668000}
\definecolor{darkorange}{HTML}{FF4500}
\definecolor{bluscuro}{HTML}{0000f5}
\definecolor{bluscuro2}{HTML}{8673a1}
\definecolor{carta}{HTML}{e0ffff}

\usepackage[export]{adjustbox}
\DeclareMathOperator{\U}{U}

\DeclareMathOperator{\SO}{SO}

\newcommand{\de}{\partial}

\newcommand{\coma}{\text{ , }}
\newcommand{\fstop}{\text{ .}}

\newcommand{\bmat}{\left(\begin{array}}
\newcommand{\emat}{\end{array}\right)}

\def\yzero{\smash{\hbox{$y\kern-4pt\raise1pt\hbox{${}^\circ$}$}}}

\def\beq{\begin{equation}}
\def\eeq{\end{equation}}
\def\beqa{\begin{eqnarray}}
\def\eeqa{\end{eqnarray}}

\def\-{\hphantom{-}}
\def\ov{\overline}
\def\s2{\frac{1}{\sqrt2}}

\def\beq{\begin{equation}}
\def\eeq{\end{equation}}
\def\beqa{\begin{eqnarray}}
\def\eeqa{\end{eqnarray}}
\def\tr{{\rm tr \,}}

\def\diag{{\rm diag \,}}
\def\IF{\relax{\rm I\kern-.18em F}}
\def\II{\relax{\rm I\kern-.18em I}}

\def\Dsl{\,\raise.15ex\hbox{/}\mkern-13.5mu D} 
\def\id{{\rm {\bf 1}}}

\def\IS{{\bf {S}}}
\def\IR{{\bf {R}}}
\def\IZ{{\bf {Z}}}
\def\IX{{\bf {X}}}
\def\IY{{\bf {Y}}}

\def\IT{{\bf {T}}}

\def\NN{{\cal {N}}}

\newcommand{\drawsquare}[2]{\hbox{%
\rule{#2pt}{#1pt}\hskip-#2pt
\rule{#1pt}{#2pt}\hskip-#1pt
\rule[#1pt]{#1pt}{#2pt}}\rule[#1pt]{#2pt}{#2pt}\hskip-#2pt
\rule{#2pt}{#1pt}}

\newcommand{\fund}{~\raisebox{-.5pt}{\drawsquare{6.5}{0.4}}~}
\newcommand{\antifund}{~\overline{\raisebox{-.5pt}{\drawsquare{6.5}{0.4}}}~}

\catcode`\@=11   
\newdimen\@rotdimen
\newbox\@rotbox  

\def\@vspec#1{\special{ps:#1}}
\def\@rotstart#1{\@vspec{gsave currentpoint currentpoint translate
   #1 neg exch neg exch translate}}
\def\@rotfinish{\@vspec{currentpoint grestore moveto}}
\def\@rotr#1{\@rotdimen=\ht#1\advance\@rotdimen by\dp#1%
   \hbox to\@rotdimen{\hskip\ht#1\vbox to\wd#1{\@rotstart{90 rotate}%
   \box#1\vss}\hss}\@rotfinish}
\def\@rotl#1{\@rotdimen=\ht#1\advance\@rotdimen by\dp#1%
   \hbox to\@rotdimen{\vbox to\wd#1{\vskip\wd#1\@rotstart{270 rotate}%
   \box#1\vss}\hss}\@rotfinish}%
\def\@rotu#1{\@rotdimen=\ht#1\advance\@rotdimen by\dp#1%
   \hbox to\wd#1{\hskip\wd#1\vbox to\@rotdimen{\vskip\@rotdimen
   \@rotstart{-1 dup scale}\box#1\vss}\hss}\@rotfinish}%
\def\@rotf#1{\hbox to\wd#1{\hskip\wd#1\@rotstart{-1 1 scale}%
   \box#1\hss}\@rotfinish}%
\def\rotate{\@ifnextchar[{\@rotate}{\@rotate[l]}}
\def\@rotate[#1]#2{\setbox\@rotbox=\hbox{#2}\@nameuse{@rot#1}\@rotbox}

\catcode`\@=12

\hypersetup{
	pdftitle={Dynamical Tadpoles and Weak Gravity Constraints},    
	pdfauthor={\textcopyright\ Alessandro Mininno, {\'A}ngel M. Uranga},     
	pdfsubject={HEP},   
	pdfcreator={pdfLaTex},   
	pdfproducer={LaTex}, 
	pdfkeywords={},
}

\begin{document}

\makeatletter
\@addtoreset{equation}{section}
\makeatother
\renewcommand{\theequation}{\thesection.\arabic{equation}}
\pagestyle{empty}
\rightline{IFT-UAM/CSIC-20-149}
\vspace{1.2cm}
\begin{center}
\LARGE{\bf Dynamical Tadpoles \\and Weak Gravity Constraints
}
\\[8mm] 

\large{Alessandro Mininno,  Angel M. Uranga\\[4mm]}
\footnotesize{Instituto de F\'{\i}sica Te\'orica IFT-UAM/CSIC,\\[-0.3em] 
C/ Nicol\'as Cabrera 13-15, 
Campus de Cantoblanco, 28049 Madrid, Spain}
\\[2mm] 
\footnotesize{ \href{mailto:alessandro.mininno@uam.es}{alessandro.mininno@uam.es} ,    \href{mailto:angel.uranga@csic.es}{angel.uranga@csic.es}}

\vspace*{6mm}

\small{\bf Abstract} 
\\[5mm]
\end{center}
\begin{center}
\begin{minipage}[h]{\textwidth}
Non-supersymmetric string models are plagued with tadpoles for dynamical fields, which signal uncanceled forces sourced by the vacuum. We argue that in certain cases, uncanceled dynamical tadpoles can lead to inconsistencies with quantum gravity, via violation of swampland constraints. We describe an explicit realization in a supersymmetric toroidal ${\textbf{Z}}_2\times{\textbf{Z}}_2$ orientifold with D7-branes, where the dynamical tadpole generated by displacement of the D7-branes off its minimum leads to violation of the axion Weak Gravity Conjecture.
In these examples, cancellation of dynamical tadpoles provides consistency conditions for the configuration, of dynamical nature (as opposed to the topological conditions of topological tadpoles, such as RR tadpole cancellation in compact spaces). We show that this approach provides a re-derivation of the Z-minimization criterion for AdS vacua giving the gravitational dual of a-maximization in 4d $\mathcal{N}=1$ toric quiver SCFTs.
\end{minipage}
\end{center}
\newpage
\setcounter{page}{1}
\pagestyle{plain}
\renewcommand{\thefootnote}{\arabic{footnote}}
\setcounter{footnote}{0}

\tableofcontents

\vspace*{1cm}

\newpage

\section{Introduction}
Supersymmetry breaking has proven a formidable challenge since the early days of string theory. Leaving aside the potential appearance of tachyons, the supersymmetry breaking ingredients often produce tadpole sources for dynamical fields, unstabilizing the vacuum \cite{Fischler:1986ci,Fischler:1986tb}. In terms of underlying supersymmetric moduli space, this can be described in terms of a non-trivial potential for the moduli, with the tadpole signaling that the theory is sitting on a slope, rather than at a minimum (or an otherwise tachyonic  extremum). Simple realizations arise in early models of supersymmetry breaking using antibranes  in type II (orientifold) compactifications \cite{Sugimoto:1999tx,Antoniadis:1999xk,Aldazabal:1999jr,Uranga:1999ib}. %
As in these models such tadpoles arise for fields in the NSNS sector, they are usually known as NSNS tadpoles. However, since similar phenomena arise in more general contexts e.g. for open string moduli (or in other constructions like heterotic or M-theory), we refer to them as {\em dynamical tadpoles}.

They are in contrast with non-dynamical tadpoles, i.e. tadpoles for non-propagating $p$-form fields (such as the familiar RR tadpoles), which  lead to topological consistency conditions on string vacua. Instead, dynamical tadpoles indicate not an inconsistency of the theory, but the fact that equations of motion are not obeyed in the proposed configuration, which should be modified to a spacetime dependent solution, e.g. rolling down the slope of the potential (see e.g. \cite{Dudas:2000ff,Dudas:2004nd,Mourad:2016xbk} for this approach in the above context). Hence, they are often treated more lightly, or directly ignored/hidden under the rug.

In this work we argue that such a mistreatment of dynamical tadpoles has a dramatic impact on the consistency of the theory, and in particular can lead to stark contradiction with Quantum Gravity,  in the form of violations of some of the best established swampland constraints \cite{Vafa:2005ui,Ooguri:2006in} (see \cite{Brennan:2017rbf,Palti:2019pca} for reviews), in particular the Weak Gravity Conjecture (WGC) \cite{ArkaniHamed:2006dz}. 

We illustrate these ideas in an explicit example of a type IIB orientifold compactification with NSNS and RR 3-form fluxes \cite{Dasgupta:1999ss,Giddings:2001yu}, with D7-branes, and admitting a supersymmetric minimum. We focus on supersymmetric instantons given by euclidean D3-branes (ED3-branes) wrapped on 4-cycles, satisfying the axion WGC \cite{ArkaniHamed:2006dz}, and in fact saturating it as the BPS relation \cite{Ooguri:2016pdq}. We consider toroidal models (and orbifolds thereof), on which D7-branes have position `moduli' which are in fact stabilized by the fluxes \cite{Gorlich:2004qm,Camara:2004jj,Gomis:2005wc,Bielleman:2015lka}. The potential arises by the axion monodromy mechanism \cite{Silverstein:2008sg,Kaloper:2008fb,Marchesano:2014mla,Hebecker:2014eua}, with the axion played by the periodic D7-brane position. Moving the D7-branes slightly off this minimum leads to a controlled supersymmetry breaking due to flux-induced extra tension on the D7-brane worldvolume, and the generation of dynamical tadpoles, in particular for the D7-brane position `modulus' itself. This kind of displacement has been exploited in the construction of models of inflation \cite{Ibanez:2014kia,Ibanez:2014swa}. 

The key point is that the flux-induced extra energy density stored on the D7-brane worldvolume sources corrections to the geometry, which are usually encoded in a corrected internal warp factor (see \cite{Kim:2018vgz}, based on \cite{Baumann:2006th} in the supersymmetric setup). We show that this procedure implies brooming a dynamical tadpole under the rug, and that it leads to a contradiction with Quantum Gravity; concretely, it produces corrections to the action of ED3-brane instantons which violate the axion WGC. 

The problem lies in the assumption that the backreaction of the supersymmetry breaking source is fully encoded in an internal warp factor, with no effect on the non-compact spacetime configuration, hence ignoring the dynamical tadpole sourced by supersymmetry breaking. Quantum Gravity is thus reminding us that consistent configurations must necessarily include spacetime dependence to account the dynamical tadpole. 

Although we showed this for a concrete model, we expect the general ideas to hold in more general configurations, and even genuinely non-supersymmetric vacua. In fact, we advocate that these general ideas must play a crucial role in understanding how swampland constraints on vacua extend to `moduli' spaces with non-trivial scalar potential. In particular, unless a proper treatment of the dynamical tadpoles is implemented, the familiar formulation of swampland constraints such as the WGC can be expected to hold {\em only} in vacua.

One can turn this logic around and consider that the condition to satisfy the WGC in its familiar formulation can be equivalent to the condition to sit at a vacuum, i.e. minimizing the corresponding scalar potential. This is indeed what happens in our D7-brane case study, and we expect this to hold in more generality.\footnote{This view aligns with the recent progress in relating swampland constraints on spacetime configurations and on properties of states defined on them, see for instance \cite{Herraez:2020tih,Lanza:2020qmt}.} In fact, we provide extra support for this idea in an amusingly unrelated setup; we show that the condition of $Z$-minimization \cite{Martelli:2005tp,Martelli:2006yb}, which in holographic dualities provides the gravitational dual of $a$-maximization  of 4d $\NN=1$ SCFTs \cite{Intriligator:2003jj} in terms of type IIB AdS$_5\times \IX_5$ vacua, follows from applying the WGC for D3-branes wrapped on 3-cycles of the internal variety $\IX_5$. 

The paper is organized as follows. In Section \ref{sec:backreactions} we consider D-brane backreaction effects, both in the supersymmetric (Section \ref{sec:susy-backreactions}) and the non-supersymmetric setups (Section \ref{sec:nonsusy-backreactions}). In Section \ref{sec:the-meat}, we describe a class of  models and discuss how mistreatment of its dynamical tadpole can lead to naive violations of the WGC. In Section \ref{sec:model} we describe the supersymmetric toroidal orientifolds with D7-branes and 3-form fluxes. In Section \ref{sec:the-tadpole} we discuss the dynamical tadpole generated when D7-branes move off its minimum, and discuss its backreaction. In Section \ref{sec:the-clash} we argue that computation of the backreaction only in the internal space can lead to violation of the WGC for different classes of ED3-brane instantons. In Section \ref{sec:the-orientifold} we construct and explicit orientifold of $\IT^6/(\IZ_2\times\IZ_2)$ realizing this idea. In Sections \ref{sec:discrete-torsion} and \ref{sec:vector-structure} we review different discrete choices in toroidal orientifolds, and in Section \ref{sec:the-model} we employ them to build our explicit example, which is equipped with suitable fluxes in Section \ref{sec:the-fluxes}. We conclude with some final considerations in Section \ref{sec:discussion}. In Appendix \ref{app:z-minimization} we illustrate that the same ideas underlie the condition of Z-minimization in AdS$_5$ vacua providing gravitational duals to large classes of 4d $\NN=1$ quiver SCFTs.

\section{D-brane backreactions in local models}
\label{sec:backreactions}

In our examples, the dynamical tadpole is sourced by D-branes, hence its proper discussion requires accounting for the D-brane backreaction. In this section we consider several examples of backreaction of D-branes on other D-branes, in cases where they preserve a common supersymmetry, or not. For our examples in later sections we need to focus on backreaction effects on ED3's, so we restrict to this case in this section, although the ideas generalize easily to other branes. We also mainly focus on local models, leaving the discussion of global models, and the issues of the resulting dynamical tadpoles, to Section \ref{sec:the-meat}.

\subsection{Warm up: Supersymmetric backreactions}
\label{sec:susy-backreactions}

We start with a review of supergravity backgrounds sourced by D-branes (D3-branes, D7-branes, and bound states thereof), and the backreaction effects on ED3-branes preserving a common supersymmetry. In these supersymmetric cases, the discussion in this section is  related to alternative description in terms of generalized calibrations, see \cite{Gutowski:1999iu,Gutowski:1999tu,Gauntlett:2001ur,Gauntlett:2002sc,Gauntlett:2003cy,Martelli:2003ki,DallAgata:2003txk,Cascales:2004qp}.

\smallskip

\subsubsection{D7 on ED3}
\label{sec:d7-on-ed3}

Consider type IIB theory on $M_4\times \IX_4\times \IR^2$, with $\IX_4$ a compact K3 or $\IT^4$, and consider $N_{\rm D7}$ D7-branes spanning the directions 01234567 and transverse to 89 (see below for remarks on the bounds on $N_{\rm D7}$). In this theory, there is a BPS instanton given by an euclidean D3-branes (ED3) spanning $\IX_4$ and localized in 0123 and 89, in general not coinciding with the D7-brane in those coordinates. The fact that the ED3 is BPS is easy to verify from the common solutions to the supersymmetry conditions of the D7 and the ED3. In an alternative view, there is a `no-force' condition, which is easy to check from the open string perspective, from the vanishing of the 1-loop annulus amplitude. Here we are instead interested in the closed string perspective, in which we check that the supergravity background created by the D7-brane exerts `no net force' on the ED3 (in more proper language, the action of the ED3 is independent of its position with respect to the D7).

We consider the background created by the D7-branes. Denoting by $z$ the complex plane in 89, and using $r=|z|$, the metric has the general brane solution structure
\beqa
ds^2\, =\, Z(r)^{-\frac 12} \eta_{\mu\nu} \,dx^\mu\, dx^\nu\, +\, Z(r)^{- \frac 12} ds_{\IX_4}^{\,2}\, +\, Z(r)^{ \frac 12} dz\,d{\bar z}\,.
\eeqa
The function $Z(r)$ obeys the 2d Laplace equation with a point source at the origin. In the non-compact case, for $N_{\rm D7}$ D7-branes, we have
\beqa
Z\, =\, -\frac{N_{\rm D7}}{2\pi}\log (r/L)\, \, +\, \ldots\,,
\label{d7z}
\eeqa
where $L$ is a scale set by e.g. by the global compactification (see Section \ref{sec:the-meat} for a related discussion), and the dots correspond to extra contributions due to possible distant sources.
In addition, there is a non-trivial background for the dilaton 
\beqa
e^{-\phi}\,=\, Z(r)
\eeqa
and for the 10d axion. These are more easily described by combining them into the 10d complex coupling $\tau=C_0+ie^{-\phi}$. For the case in Eq. \eqref{d7z}, it is given by
\beqa
\tau\,=\, \frac{N_{\rm D7}}{2\pi i}\log (z/L)\, +\ldots
\eeqa
This encodes the shift $C_0\to C_0+1$ upon the shift $z\to e^{2\pi i} z$ as one surrounds one D7 .

This is a good approximation for a small number of D7-branes; due to the non-flat asymptotics, larger $N_{\rm D7}$ overcloses the transverse space to a compact structure better described in F-theory \cite{Vafa:1996xn}. In our description we will deal with one D7-brane and, if necessary, we consider the compact configuration close to the weakly coupled IIB limit of  the orientifold  of $\IT^2$ by $\Omega {\cal R} (-1)^{F_L}$, where ${\cal R}:z\to -z$ introducing O7-planes \cite{Sen:1996vd}. In such situation, the above function $Z$ should be replaced by a $\IT^2$ Green's function, see later for analogous examples.

It is now straightforward to consider an ED3 in the probe approximation, and to evaluate its action to check it is independent of its position relative to the D7-brane. Concretely, the effect of the backreaction is to introduce factors of $Z$ which cancel off
\beqa
S_{\rm ED3}\, =\, \frac{\left(Z^{-\frac 14}\right)^4{\rm Vol}(\IX_4)}{Z^{-1} g_s}=\, \frac{{\rm Vol}(\IX_4)}{g_s}\,.
\eeqa
This is the closed string version of the BPS property.

As usual, the open string description is more suitable for inter-brane distances below the string scale, while the closed string exchange description is better suited for larger distances. In this case the discussion is equivalent in both pictures, due to the large amount of supersymmetry in the system.

\subsubsection{D3 on ED3}
\label{sec:d3-on-ed3}

Les us now consider a different example. Consider again type IIB theory on $M_4\times \IX_4\times \IR^2$ as above, with $\IX_4$ compact, and consider $N_{\rm D3}$ D3-branes along  4d Minkowski space. Although the discussion can be carried out in more generality, we are interested in smearing the D3-branes as a constant density along $\IX_4$. The backreaction thus depends only on the radial direction  in the complex plane $z=re^{i\theta}$ spanned by 89. This is similar to the above D7-brane case, and in fact both are related by `T-duality'\footnote{This would be Fourier-Mukai in case $\IX_4=$K3.} in $\IX_4$. Adapting the celebrated D3-brane supergravity solution to the case of a warp factor obeying a 2d Laplace equation, we have
\beqa
ds^2 & = & Z(r)^{-\frac 12} \eta_{\mu\nu} \,dx^\mu\, dx^\nu\, +\, Z(r)^{\frac 12} ds_{\IX_4}^{\,2}\, +\, Z(r)^{ \frac 12} dz\,d{\bar z}\,,
\label{metric-d3}
\eeqa
with
\beqa
Z(r)\,=\, -\frac {g_s N_{\rm D3} }{2\pi} \log (r/L) \, +\,\ldots
\eeqa
Here the dots denote extra pieces, due to global structure e.g. due to possible distant sources, and $L$ again denotes a global (e.g. compactification) scale.

The D3-branes also source the RR 4-form $C_4$. Given its self-duality, the background can be expressed in terms of the components of $C_4$ along $\IX_4$. This leads to the following profile with the polar angle $\theta$ 
\beqa
\varphi\, \equiv\, \int_{\IX_4} C_4\,  =\, \frac {N_{\rm D3}}{2\pi} \, \theta +\ldots
\label{4form-axion-D3}
\eeqa

We now consider a BPS instanton given by an ED3 wrapped on $\IX_4$, and describe the effect of the backreaction. The ED3 feels the warping in the metric, and couples directly to the axion $\varphi$, so its Dirac-Born-Infeld + Chern-Simons action picks up a factor
\beqa
\frac{1}{g_s}\, \left(\, -\frac {g_s N_{\rm D3}}{2\pi}  \, \log r\, \right) \, -\, i\,\frac {N_{\rm D3}}{2\pi}\, {\rm Im}\  \log z \,+\, \ldots =\, -\frac{ N_{\rm D3}}{2\pi}\, \log z\, + \,\ldots 
\label{d3oned3-D3}
\eeqa
In this case, it is the holomorphy of the result that encodes the BPS nature of the ED3.
Our computation is essentially that in \cite{Baumann:2006th}; indeed, using \eqref{d3oned3-D3}  as the corrected instanton action, for e.g. $N_{\rm D3}=1$ the 4d non-perturbative contribution to e.g. the superpotential gives
\beqa
W\, =\, z\, e^{-S_{\rm ED3}^0}\,,
\eeqa
where the factor $2\pi$ has been reabsorbed and $S_{\rm ED3}^0$ is the instanton action in the absence of correction. For the open string perspective on this result, see \cite{Ganor:1996pe,Blumenhagen:2006xt,Ibanez:2007rs,Florea:2006si}.

\subsubsection{\bf D7/D3 on ED3}
\label{sec:d7d3-on-ed3}

By combining the results of the previous two sections, it is straightforward to study the backreaction of BPS bound states of D7- and D3-branes (namely, magnetized D7-branes \cite{Angelantonj:2000hi,Blumenhagen:2000wh,Aldazabal:2000dg}). The gravitational backreaction is obtained using the harmonic superposition rule in supergravity \cite{Ortin:2015hya}
\beqa
ds^2 & = & Z_{\rm D7}^{-\frac 12}\,Z_{\rm D3}^{-\frac 12} \,\eta_{\mu\nu} \,dx^\mu\, dx^\nu\, +\, Z_{\rm D7}^{- \frac 12}\, Z_{\rm D3}^{\frac 12} \,ds_{\IX^4}^{\,2}\, +\, Z_{\rm D7}^{ \frac 12}\,Z_{\rm D3}^{ \frac 12} \,dz\,d{\bar z} \,,
\eeqa
with
\beqa
Z_{\rm D7}\,=\, -\frac {N_{\rm D7} }{2\pi} \log (r/L) \quad ,\quad Z_{\rm D3}\,=\, -\frac {g_s N_{\rm D3} }{2\pi} \log (r/L)\,,
\eeqa
where we are ignoring the dots. In addition, we have backgrounds for the IIB complex coupling $\tau$ and the axion $\varphi$ as in Eq. \eqref{4form-axion-D3}
\beqa
\tau\,=\, \frac{N_{\rm D7}}{2\pi i}\log (z/L) \quad ,\quad \varphi \,  =\, \frac {N_{\rm D3}}{2\pi} \,{\rm Im}\, (\,\log z)\,.
\eeqa
In the ED3 action, the dilaton background cancels with the D7-brane metric backreaction factor $Z_{\rm D7}$, leaving a net effect due only to the D3-branes, given by \eqref{d3oned3-D3}.

\medskip

The generalization of results of the previous sections to global supersymmetric setups is clear, by simply replacing $Z_{\rm D7}$, $Z_{\rm D3}$ by the corresponding solutions of the Laplace equation
\beqa
-\bigtriangleup \, Z_{\rm D3/7}\left(x^8,x^9\right)\, =\, \rho_{\rm D3/7}\left(x^8,x^9\right)\,.
\label{laplace}
\eeqa
The above examples correspond to the local solutions of $\rho_{\rm D3/7}\sim N_{\rm D3/7}\,\delta_2(z,{\bar z})$. In global compact setups, the Laplace equation implies some non-trivial integrability conditions on the sources, which are closely related to the dynamical tadpoles. We thus postpone their discussion to Section \ref{sec:the-meat}.

\subsection{Non-supersymmetric backreactions}
\label{sec:nonsusy-backreactions}

In this section we consider supersymmetry breaking sources, including anti-D3-branes (dubbed ${\overline{\rm D3}}$-branes in the following), and their backreaction effect on ED3's. Our aim is to obtain the backreaction of D7-branes with induced D3/${\overline{\rm D3}}$-brane charge, as arises in the presence of NSNS 2-form field backgrounds (ubiquitous in compactifications with NSNS and RR 3-form fluxes \cite{Dasgupta:1999ss,Giddings:2001yu}). Our results provide a simple re-derivation of \cite{Kim:2018vgz} (to which we refer the reader for a detailed discussion), and generalize easily to some further effects not considered therein.

\subsubsection{Anti-D3 on ED3}
\label{sec:antid3-on-ed3}

Consider the setup  of type IIB theory on $M_4\times \IX_4\times \IR^2$ as in Section \ref{sec:d3-on-ed3}, but with $N_{\overline{\rm D3}}$ ${\overline{\rm D3}}$-branes instead of D3-branes. We consider the model locally, so that we ignore global tadpoles. Since the ${\overline{\rm D3}}$-branes have the same tension but opposite charge compared with D3-branes, their backreaction on the metric is given by \eqref{metric-d3} with
\beqa
Z(r)\,=\, -\frac {g_s N_{\ov{\rm D3}} }{2\pi} \log (r/L) \, +\,\ldots
\eeqa
and the backreaction on the RR 4-form equals to Eq. \eqref{4form-axion-D3} with an extra sign
\beqa
\varphi\, =\, - \frac {N_{\ov{\rm D3}}}{2\pi} \, \theta +\ldots\,.
\label{4form-axion-bD3}
\eeqa
The effect on an ED3 located at position $z$ is multiplication by a factor
\beqa
-\frac{ N_{\rm D3}}{2\pi}\, \log ({\bar z}/L)\, + \,\ldots \,.
\eeqa
This anti-holomorphic dependence reflects the fact that locally, the ED3 and ${\overline{\rm D3}}$ preserve common supersymmetries, albeit those opposite to the ED3/D3 system (and globally, by the CY threefold compactification).

\subsubsection{D7/D3/anti-D3 on ED3}
\label{sec:d7d3antid3-on-ed3}

We could now consider the backreaction of D3/${\overline{\rm D3}}$ pairs. However, these systems are strongly unstable due to tachyons, and we prefer to consider a more tractable alternative, which in fact is our main setup in future sections. We consider type IIB theory on $M_4\times \IX_4\times \IR^2$ with a D7-brane wrapped on $\IX_4$ with equal  smeared (in $\IX_4$) D3/${\overline{\rm D3}}$ charge distributions. These arise in the presence of a worldvolume gauge background with field strength $F_2$ and/or pullbacked NSNS 2-form background $B_2$, which combine into
\beqa
{\cal F}_2\,=\, 2\pi \alpha'\, F_2\, +\, B_2\,.
\label{curly-f}
\eeqa
The D3/${\overline{\rm D3}}$ charge distributions cancel locally when it satisfies 
\beqa
{\cal F}_2\wedge {\cal F}_2=0\,.
\label{no-charge-condition}
\eeqa 
The individual D3- and ${\overline{\rm D3}}$-brane contributions are obtained by extracting the self- and anti-selfdual pieces 
\beqa
{\cal F}_{2,\pm}=\frac 12 ({\cal F}_{2}\pm \star_{4} {\cal F}_{2})\,,
\eeqa
where $\star_4$ is Hodge in $\IX_4$. In particular, we have
\beqa
N_{\rm D3}\, =\, \int_{\IX_4} {\cal F}_{2,+}\wedge {\cal F}_{2,+}\quad , \quad N_{\ov{\rm D3}}\, =\,- \int_{\IX_4} {\cal F}_{2,-}\wedge {\cal F}_{2,-}
\eeqa
and \eqref{no-charge-condition} implies $N_{\rm D3}=N_{\ov{\rm D3}}$, as anticipated.

Since both contribute in the same way to the gravitational background, it is useful to introduce
\beqa
N_{\rm 3}\, =\, \int_{\IX_4} |{\cal F}_{2}|^2 \, =\, N_{\rm D3}\,+\, N_{\ov{\rm D3}}\, = \, 2N_{\rm D3}\,.
\label{n3}
\eeqa

To compute the backreaction, we superimpose the effects of the corresponding brane charges, as computed in earlier sections. This is the leading contribution in an expansion with the sources (flux densities) as perturbative parameter; this implies ignoring corrections that would involve e.g. solving the gravitational background of a source in the background created by another source. The present expansion fits well with the regime needed for coming sections, and agrees with the detailed analysis in \cite{Kim:2018vgz}.

The result is that the correction to the ED3 action is controlled by a factor $-\frac{ N_3}{2\pi}\, \log (r/L)$. Particularizing to toroidal $\IX_4=\IT^4$ and constant backgrounds, we have
\beqa
S_{\rm ED3}\,  =\, \left[\,1\,-\,\frac 1{2\pi} |{\cal F}_{2}|^2\, \log (r/L)\, + \,\ldots\,\right]\, S_{\rm ED3}^0\,,
\label{d3oned3}
\eeqa
where  for future convenience we have added a constant piece in the prefactor, so that the action is $S_{\rm ED3}^0$ when ${\cal F}_{2}=0$.
The above result is easily understood from different perspectives. The D7-brane backreaction on the dilaton and metric cancel out, leaving an effective D3- and ${\overline{\rm D3}}$-brane distribution, whose backreaction on $C_4$ cancels exactly, and whose backreaction on the metric add up. As anticipated, we recover the result in \cite{Kim:2018vgz}.

As a final remark, we emphasize that, although the configuration is non-supersymmetric, it does not lead to tachyons. Indeed open strings with both endpoints on the D7-brane are insensitive to the worldvolume magnetic flux, and are hence tachyon-free. On the other hand, in forthcoming compact models with orientifolds, open strings between the D7-brane and its orientifold image have stretching contributions to their squared mass which overcome the potential tachyonic contributions from the non-supersymmetric flux. In fact, in the long inter-brane distance regime, the non-supersymmetric open string sector leads, in the closed string channel interpretation, to the backreaction interactions just discussed. Similar remarks apply to examples in coming sections.

\subsubsection{D5/anti-D5 on ED3}
\label{sec:d5antid5-on-ed3}

We now consider an effect present in this setup, but not included in \cite{Kim:2018vgz}. In the presence of the worldvolume background ${\cal F}_{2}$, there is an induced D5- or ${\ov{\rm D5}}$-brane density, which also must backreact on the geometry. For simplicity and future use, we consider $\IX_4=\IT^4$ (or an orbifold thereof), and consider constant ${\cal F}_{2}$, so that the D5-brane charge is smeared, and the solution does not depend on internal coordinates in $\IX_4$. We also focus on induced D5-brane charge, and the results will extend easily to D5/${\ov{\rm D5}}$ setups needed later on.

The supergravity background created by a D5-brane (e.g. along 45 and transverse to 67 in $\IX_4$) is determined by a 2d harmonic function $Z_{\rm D5}(r)$ as
\begin{equation}
\begin{split}
ds^2 & =  Z_{\rm D5}^{-\frac 12}\, \,\eta_{\mu\nu} \,dx^\mu\, dx^\nu\, +\, Z_{\rm D5}^{- \frac 12}\,  \,ds_{45}^{\,2}\, +\, Z_{\rm D5}^{ \frac 12}\,  ds_{67}^{\,2}\,+\, Z_{\rm D5}^{\frac 12} \,dz\,d{\bar z} \,, \\
e^{-2\phi} &= Z_{\rm D5}\,.
\end{split}
\end{equation}
In local $\IR^2$ we have
\beqa
Z_{\rm D5}\,=\, - \frac {g_s N_{{\rm D5}} }{2\pi} \log (r/L) \, +\,\ldots\,.
\label{n-five}
\eeqa
The above backreaction superimposes to D7/D3/${\ov{\rm D3}}$ as discussed earlier. 

There is also a RR 2-form background, which will not be relevant to our setups. In fact, we are interested in obtaining the correction to the action of an ED3 (along 4567) from the above background, and the latter does not couple to the RR 2-form. Note also that in this case there is no correction coming from the backreacted metric, since the $Z_{\rm D5}$ factors have inverse power for 45 and 67, and they cancel off. One is thus left with the correction to the dilaton, which gives a correction to $S_{\rm ED3}^0$ by a factor
\beqa
S_{\rm ED3}^0\, \to \,\left[\, 1\, -\frac 1{4\pi} |{\cal F}_{2}|\, \log (r/L)\, + \,\ldots\,\right]\, S_{\rm ED3}^0\,,
\eeqa
where we expanded $Z_{\rm D5}^{\frac 12}$ to first order in the induced D5-brane charge, and we already included the constant piece 1, as above.

In the above expressions we have been sloppy concerning numerical factors, which are not essential to our analysis below, since it is enough to keep track of the parametric dependence with induced charges. The signs of the different contributions are on the other hand crucial.

\section{Dynamical tadpoles and WGC in D7-brane models}
\label{sec:the-meat}

\subsection{A D7-brane model}
\label{sec:model}

We are considering the model with D7-brane and 3-form fluxes in \cite{Gomis:2005wc}, which we now review. Although the specific model fulfilling the conditions we need is discussed in Section \ref{sec:the-orientifold}, we here discuss the general class of type IIB orientifolds of $\IT^6$, or rather orbifolds thereof, like $\IT^2\times \IT^4/\IZ_2$, or $\IT^6/(\IZ_2\times \IZ_2)$. For concreteness, we carry out the description in terms of the latter, although we also discuss the simpler alternatives when indicated. We take a factorized $(\IT^2)^3$ structure, with coordinates $0\leq x^i,y^i\leq 1$, $i=1,2,3$ for each $\IT^2$, and complexify them as $z^i=x^i+\tau_i y^i$. We mod out by $\Omega {\cal R}(-1)^{F_L}$, where ${\cal R}$ flips all $\IT^6$ coordinates, $z^i\to -z^i$, which introduces 64 O3-planes. In addition, in the presence of orbifold quotients, e.g. $\IZ_2\times \IZ_2$, there are additional sets of 4 O7$_i$-planes localized on the i$^{\text{th}}$ $\IT^2$. The O3-plane charge is canceled against contributions from the upcoming 3-form fluxes, and D3-branes if necessary, which can be located at arbitrary positions. The models typically also contain D7$_i$-branes, transverse to the i$^{\text{th}}$ $\IT^2$. These can be located at arbitrary positions, provided we include the corresponding orbifold and orientifold images, and that we comply with the flux stabilization, discussed next.

Following \cite{Gomis:2005wc} we introduce a specific choice of NSNS and RR 3-form fluxes, 
\begin{equation}
\begin{split}
F_3 &= 4\pi^2\alpha'\, N\, \left(\, dx^1\wedge dx^2\wedge dy^3\, +\, dy^1\wedge  dy^2\wedge  dy^3\,\right) \,, \\
H_3 &= 4\pi^2\alpha'\, N\, \left(\, dx^1\wedge dx^2\wedge dx^3\, +\, dy^1\wedge dy^2\wedge dx^3\,\right) \,.
\end{split}
\label{flux-def}
\end{equation}
Although naively $N\in \IZ$ due to flux quantization, it must actually be some suitable multiple of some $N_{\rm min}$ due to the diverse quotients; for instance, $N\in 2\IZ$ for $\IT^6$ orientifolds with standard (i.e. negatively charged) O3-planes \cite{Frey:2002hf}, and $N\in 4\IZ, 8\IZ$ in $\IT^6/(\IZ_2\times \IZ_2)$ orientifolds \cite{Blumenhagen:2003vr,Cascales:2003zp}, as we recall in Section \ref{sec:the-fluxes}. 

The flux superpotential admits supersymmetric minima for 
\beqa
\tau_1\tau_2=-1\quad ,\quad \tau_3\tau=-1\,,
\eeqa
where $\tau$ is the 10d IIB complex coupling.

The fluxes also stabilize some of the D7-brane moduli as follows. The presence of the fluxes introduces an in general non-zero pullback of the NSNS 2-form on the D7-branes. For instance, for a D7$_1$ at a generic position $\left(x^1,y^1\right)$, in a suitable gauge we have
\beqa
B|_{{\rm D7}_1} &= &4\pi^2\alpha'\, N\, \left(\, x^1\, dx^2\wedge dx^3\, +\, y^1\, dy^2\wedge  dx^3\,\right)\,.
\label{bfield-ond71}
\eeqa
The supersymmetry condition \cite{Marino:1999af} requires that \eqref{curly-f} is primitive, and of type $(1,1)$ when expressed in complex components. This is clearly satisfied at the origin $z^1=0$, where D7$_1$-branes can thus be stabilized. However, as emphasized in \cite{Gomis:2005wc} it is possible to locate them at other positions $\left(x^1,y^1\right)$ if $Nx^1, Ny^1\in \IZ$, by compensating \eqref{bfield-ond71} with suitably quantized worldvolume gauge fluxes. For our purposes, we just need some D7-brane to be located at the origin, and we can consider a general distribution for the rest. For concreteness, using  the above freedom, and the fact that $N$ is even from flux quantization, we choose to locate the D7$_1$-branes distributed in sets of 8 on top of the O7$_1$-planes so as to have local charge cancellation (see Section \ref{sec:the-model} for an explicit example).

For orientifolds of $\IT^6/(\IZ_2\times\IZ_2)$ there are typically\footnote{Albeit, not in the specific example to be constructed in Section \ref{sec:the-orientifold}.} additional kinds of D7-branes, that we now discuss. D7$_2$-brane behave similar to the D7$_1$-branes above, and introduce no qualitative new features. On the other hand, for D7$_3$-branes, motion away from the origin is compatible with supersymmetry and corresponds to a flat direction. This is because the induced B-field is $(1,1)$ (and primitive) hence satisfies the supersymmetry conditions. Our focus is in supersymmetry breaking effects, so we will not be interested in exploring this possibility.

Focusing again on D7$_1$-branes, we also note that, despite the induced B-fields, there is no net induced D3-brane charge, since \eqref{bfield-ond71} wedges to zero with itself, c.f. \eqref{no-charge-condition}. In fact in other models, or even in this model but for the motion of the D7$_3$-branes, there is a non-zero net induced D3-brane charge, proportional to the displacement squared. This is compatible with the cancellation of tadpoles for the RR 4-form due to a mechanism unveiled in \cite{Ibanez:2014swa}: the backreaction of the induced D5-brane charges on the D7-brane modifies the RR 3-form flux $F_3$, changing the flux contribution to the tadpole in precisely the right amount. In our examples we focus on D7-brane motions not involving induced net D3-brane charge, and hence no such modification of the $F_3$ flux background.

\subsection{Moving off the minimum and the dynamical tadpole}
\label{sec:the-tadpole}

We  now start addressing our main point, by introducing a source of supersymmetry breaking, which triggers a dynamical tadpole. A simple way to achieve this is to consider moving away from the minimum of the potential. Among the different ways to climb up the potential, we focus on the motion of D7$_1$-branes because they lead to better understood backreaction effects, of the kind discussed in Section \ref{sec:backreactions} (clearly, D7$_2$-brane motion leads to similar results).

Consider the above type IIB orientifold, and consider a fixed point at which some D7$_1$-branes sit, which without loss of generality we take to be the origin $x^1=y^1=0$. To be precise, we consider regular D7-branes with respect to the relevant orbifold group, so that they can move off into the bulk (see Section \ref{sec:the-orientifold} for a detailed discussion of constructions allowing this motion) e.g. in the first complex plane, to the position
\beqa
z^1\, =\, \pm \epsilon\, \in\IR\,,
\eeqa
where the two signs correspond to a D7-brane and its image. 

This motion is along a massive direction, off the minimum of the potential, due to the non-trivial B-field \eqref{bfield-ond71} on the D7$_1$-brane worldvolume. Since there is no net D3-brane charge, this can be regarded as a D3/${\overline{\rm D3}}$-brane tension localized at  $z^1=\pm \epsilon$, and proportional to $N^2|\epsilon|^2$, backreacting on the metric, as in Section \ref{sec:antid3-on-ed3}. This is also the scaling of the potential energy stored in the configuration. In addition, there is an induced D5-brane change on the D7-brane (and its corresponding ${\overline{\rm D5}}$-brane in its image), which imply a backreaction on the dilaton, as in Section \ref{sec:d5antid5-on-ed3}. Notice, that this is a consequence of the fact the the induced D5-  and ${\overline{\rm D5}}$-branes sit at different locations (related by the orientifold action) in the internal space, hence can lead to a non-trivial backreaction on different fields, even if the total brane charge adds up to zero in the internal space, as demanded by RR tadpole cancellation (as there are no O5-planes) or relatedly, by the fact that the (zero mode of the) corresponding RR field is projected out by the orientifold. We are now interested in computing the backreaction of these extra sources. Since everything will be happening in a complex plane, from now on we denote $z^1$ and $\tau_1$ by $z$ and $\tau$ (hoping for no confusion with the IIB complex coupling).

Since we consider a motion $|\epsilon|\ll L$, where $L$ sets the size of the $(\IT^2)_1$ directions, we can start with a local model as in Section \ref{sec:backreactions}. Note that we still consider $M_s\ll\epsilon$ so that we can use the supergravity description to obtain the backreaction. In general, there is a non-trivial supergravity background created by the D3-branes and O3-planes in the configuration, the D7-branes and O7-planes, and finally the induced D3/${\overline{\rm D3}}$-branes, and D5- (and ${\overline{\rm D5}}$-) branes. In coming sections we are interested in the effect of this background on ED3$_1$-branes,\footnote{In Section \ref{sec:the-fractional} we also consider
ED$(-1)$-brane instantons. These are also BPS with respect to the supersymmetric D3/O3 and D7/O7 background, and again disappear in the relevant part of the backreaction.} for which most of these contributions cancel. As discussed in Section \ref{sec:nonsusy-backreactions}, the key backreaction effects are in the warp factor $Z$ sourced only by the induced tension on the D7-brane worldvolume, and the induced D5-brane backreaction on the dilaton. They schematically read
\begin{equation}
\begin{split}
Z&\simeq\, 1-\frac{N^2|\epsilon|^2}{2\pi}\, \left[ \, \log \, (\,|z-\epsilon|/L)\, +\,\log \, (\,|z+\epsilon|/L)\,  \right]+ \ldots \\
g_s^{-1} &\simeq\, 1\, -\frac {N|\epsilon|}{4\pi} \,\left[ \, \log \, (\,|z-\epsilon|/L)\, +\,\log \, (\,|z+\epsilon|/L)\,  \right]+ \ldots
\label{full-local-Z}
\end{split}
\end{equation}
where we introduce the constant term 1 to recover trivial backreaction for $\epsilon=0$. Here the overall prefactors depending on $N|\epsilon|$ provide the induced brane density, and the dots hide global features to which  turn next. Note that in these and coming expressions, we are only interested in the parametric dependence and we skip order 1 numerical factors, in particular in the coefficients of the log terms for the metric and dilaton profiles. On the other hand, the explicit minus sign and the structure of the bracket of logs itself is identical for both backgrounds, as it is determined by the solution of a Laplace equation in $\IR^2$ with sources at $z=\pm \epsilon$.

The above expression is valid for small $\epsilon$, since we take the linear/quadratic approximation for the induced D5/D3-brane density. The result can however be extended for larger $\epsilon$ by using the full DBI contribution, e.g. for the 3-brane charge we sketchily change
\beqa
|\epsilon|^2\, \rightarrow\, 2\left(\sqrt{1+|\epsilon|^2} -1\right)\,.
\eeqa
We will nevertheless stick to small $\epsilon$, since it controls the expansion of weak supersymmetry breaking sources employed in the computation of the leading backreaction effect. In any event, the only relevant information is that the coefficients of the logs are positive definite (up to the explicitly indicated sign) for any value of $\epsilon$, and vanish at $\epsilon=0$ at least as $\cal{O}(\epsilon)$. 

\smallskip

We now turn to a very important point. We eventually need the backreaction at a general position $z$, not necessarily close to the origin. Hence, even if $\epsilon$ is small, we need to consider the global compactification. For this, in principle, one would simply promote the previous logarithmic backreaction to a solution of the Laplace equation with a delta function source c.f. \eqref{laplace}. However, this leads to a problem of integrability of the equation, as the left hand side integrates to zero in a compact space, and the right hand side does not. This is nothing but the dynamical tadpole problem presented in the introduction: there is a vacuum energy stored in the internal space, which leads to an inconsistency of the equations of motion.

An usual procedure (see \cite{Kim:2018vgz}, based on \cite{Baumann:2006th} in the supersymmetric setup) is to modify the equations of motion (the Laplace equation) by introducing a constant distribution of background source compensating the delta function (i.e. so that the right hand side integrates to zero). In other words, we promote
\begin{equation}
\log \left(\frac{|z+\epsilon|}{L}\right)\, +\,\log \left(\frac{|z-\epsilon|}{L}\right) \,\rightarrow\, 
G_2\left(\frac{|z+\epsilon|}{L}\right)\, +\, G_2\left(\frac{|z-\epsilon|}{L}\right)\,,
\label{promotion}
\end{equation}
where $G_2(z)$ is the 2d Green's function, satisfying
\beqa
\bigtriangleup G_2\left(z-z'\right)\,= \, \delta_2\left(z-z'\right)\, -\, \frac{1}{L^2\,{\rm Im}\,\tau}\,.
\label{the-trick}
\eeqa
Here $L$ is explicitly the length of the $\IT^2$ sides, set to $L=1$ in what follows for simplicity. The solution is given by (see e.g. \cite{Annals,Kim:2018vgz}, also \cite{Andriot:2019hay} in a different context)
\beqa
G_2(z)\, =\, \frac{1}{2\pi} \,\log \left| \frac{\vartheta_1(z| \tau)}{ \eta(\tau)} \right|\, - \frac{({\rm Im}\, z)^2}{2\,{\rm Im}\,\tau} \,,
\label{eq:Greenfunc}
\eeqa
with, defining the nome $q=e^{2\pi i \tau}$,
\beqa
\begin{split}
	\vartheta_1(z|\tau)&=2q^{1/8}\sin(\pi z)\prod_{m=1}^\infty \left(1-q^m\right)\left(1-2\cos(2\pi z)q^m+q^{2m}\right)\,,\\
	\eta(\tau)&=q^{1/24}\prod_{m=1}^\infty (1-q^m)\,,
\end{split}
\eeqa
and where an additive integration constant has been fixed to have the Green function integrated to zero (so that a constant density of source gives rise to no correction).

\smallskip

The above trick is a well-defined mathematical procedure, but its physical meaning is questionable.  It corresponds introducing by hand in \eqref{the-trick} a constant negative tension background in the internal geometry, which indeed sounds troublesome. Alternatively, it corresponds to ignoring the dynamical tadpole (potential for the D7-brane position off its minimum) and to insist that the configuration still admits a solution with distortions only in the internal space, keeping the external 4d Minkowski spacetime. In the following section we argue that these are not just subtle technicalities, but, rather that putting the dynamical tadpole under the rug, can lead to direct contradiction with quantum gravitational swampland constraints, in particular the Weak Gravity conjecture.

\subsection{The clash with the WGC}
\label{sec:the-clash}

In this section we show that in the theory there are objects that implement the WGC in the vacuum, but for which the above discussed backreaction (with the tadpole hidden under the rug) induces corrections in the {\em wrong} direction, so that the configuration no longer obeys the WGC.

A simple possibility is to focus on BPS objects at the supersymmetric minimum, and to track their properties in the displaced configuration. As anticipated in Section \ref{sec:nonsusy-backreactions}, we  consider the axion WGC and focus on the ED3-brane instantons wrapped on $\IX_4$, and transverse to the $z^1$-complex plane. We consider two possibilities, to be discussed in turn, regular ED3-branes, which are mobile and can be located at different positions in $z^1$, and fractional ED3-branes, which are stuck at a given fixed point (and can be regarded as ED3/ED$(-1)$ bound states).

\subsubsection{The regular ED3}
\label{sec:regular-ED3}

Consider a regular ED3, namely one that can be located at any position in $z^1$, before the introduction of fluxes (since the latter lead to localization, as will be mentioned soon). In the case of a $\IT^6$ orientifold, this is achieved by introducing an orientifold image, and if there are orbifold quotients this requires a specific choice of Chan-Paton actions, whose details we skip (see Section \ref{sec:the-orientifold} for an extensive discussion).

At the supersymmetric minimum, the action for the BPS ED3-brane instanton is 
\beqa
S_0\,=\, {\rm Im}\, T\,,
\eeqa
where $T$ is the 4-cycle modulus of the underlying $\IT^6$. The BPS relation ensures these instantons saturate the axion WGC, in fact also for arbitrary instanton charge (e.g. multiply wrapped ED3s).

Although the above observations apply for instantons located at arbitrary $z^1$, the introduction of 3-form fluxes leads to a localization effect, since, e.g. away from $z^1=0$ the ED3 picks up a B-field exactly as in \eqref{bfield-ond71}, which contributes to increase its action. This contribution grows quadratically with $|z^1|$ and is not suppressed for small $\epsilon$, so it dominates over $\epsilon$-dependent backreaction corrections. Hence in the following we focus on the effect of backreactions on ED3 located at $z^1=0$, where the direct flux-induced localization vanishes (however, we advance that ED3s at other possible locations will bite back in Section \ref{sec:the-landscape}). 

When we move the D7-brane off its minimum, the backreaction \eqref{full-local-Z} enters as a multiplicative factor in the instanton action. Hence, the correction to the ED3 action at $z=0$ is 
\begin{equation}
\Delta S\, =\, -\,C  \left(\,\log \left|\epsilon\right|\,\right)\, {\rm Im}\, T\,,
\label{local-ed3-correction}
\end{equation}
with a positive definite coefficient $C$, vanishing for $\epsilon\to 0$, with the sketchy structure
\beqa
C\,\sim\,\frac {N|\epsilon|}{2\pi}\,+\,\frac{N^2|\epsilon|^2}{\pi}\,.
\eeqa
Note that $N|\epsilon|$, $N^2|\epsilon|^2 {\rm Im}\,T$ encodes the total induced 5-brane and 3-brane tensions, and correspond to $N_5$, $N_3$ in \eqref{n-five}, \eqref{n3}, respectively.

Since we have small $\epsilon$, this gives a positive correction $\Delta S$ positive, thus increasing the action of the instanton. Actually this is not yet problematic. In fact, since \eqref{local-ed3-correction} follows from just the local model approximation, there is no finite 4d  Planck scales and hence no contradiction with the WGC so far. Notice also that the dynamical tadpole has not yet been ignored, since the Laplace equation in non-compact spaces does not require the cancellation of sources, hence the introduction of the fake background to cancel the tadpole. This agrees with our picture that it is the mistreatment of the dynamical tadpole that leads to problems with quantum gravitational constraints.

\subsubsection{Going compact: The ED3 landscape}
\label{sec:the-landscape}

In this section we consider the compact $\IT^2$ model, and the modification it implies for the ED3-brane action and its interplay with the axion WGC. 

As we have explained, to describe the backreaction in the compact $\IT^2$ we have to promote the logs in earlier expressions to solutions of the 2d Laplace equation c.f. \eqref{promotion}, which leads to the dynamical tadpole problem. Getting rid of it as described above, the effect of the backreaction on the action of an ED3 at a general location $z$ is
\beqa
-\,C G_{\rm tot.}(z;\epsilon)\,\equiv -C \left[\, G_2(z+\epsilon|\tau)\, +\, G_2(x-\epsilon|\tau)\,\right]\,,
\label{g-tot}
\eeqa
where $G_{\rm tot.}$ is the total Green's function due to the two sources at $z=\pm\epsilon$.
This by itself does not lead to a substantial modification, since its local behavior near $z=0$ is just the above logs, so we recover the same correction to the action of the ED3 at $z=0$. However, there is an important novelty in the compact model, since there are locations away from $z=0$ where the B-field induced on the volume of the ED3-brane can be canceled by a suitable worldvolume magnetic flux, giving a vanishing \eqref{curly-f} on the ED3. Namely, recalling \eqref{bfield-ond71}, we see that for an ED3 at $x^1=n_1/N$, $y^1=n_2/N$, with $n_i\in\IZ$, we may cancel the induced B-field by choosing
\beqa
F_2 &= &- \left(\, n_1\, dx^2\wedge dx^3\, +\, n_2 \, dy^2\wedge dx^3\,\right)\,.
\eeqa
This is nothing but the ED3 version of the open string landscape of \cite{Gomis:2005wc}. These ED3s are in the same topological charge sector as the original ED3 at $z=0$ because $\int_{\IX_4}F_2\wedge F_2=0$. Hence, the condition that we actually obtain a violation of the WGC is that the backreacted ED3 action increases for all points of the ED3 open string landscape. In terms of \eqref{g-tot}, the condition is that the value of $G_{\rm tot.}$ is negative at {\em all} the ED3 open string landscape points.

\begin{figure}[h!]
		\centering
		\begin{subfigure}{.5\textwidth}
			\includegraphics[width=\textwidth,keepaspectratio]{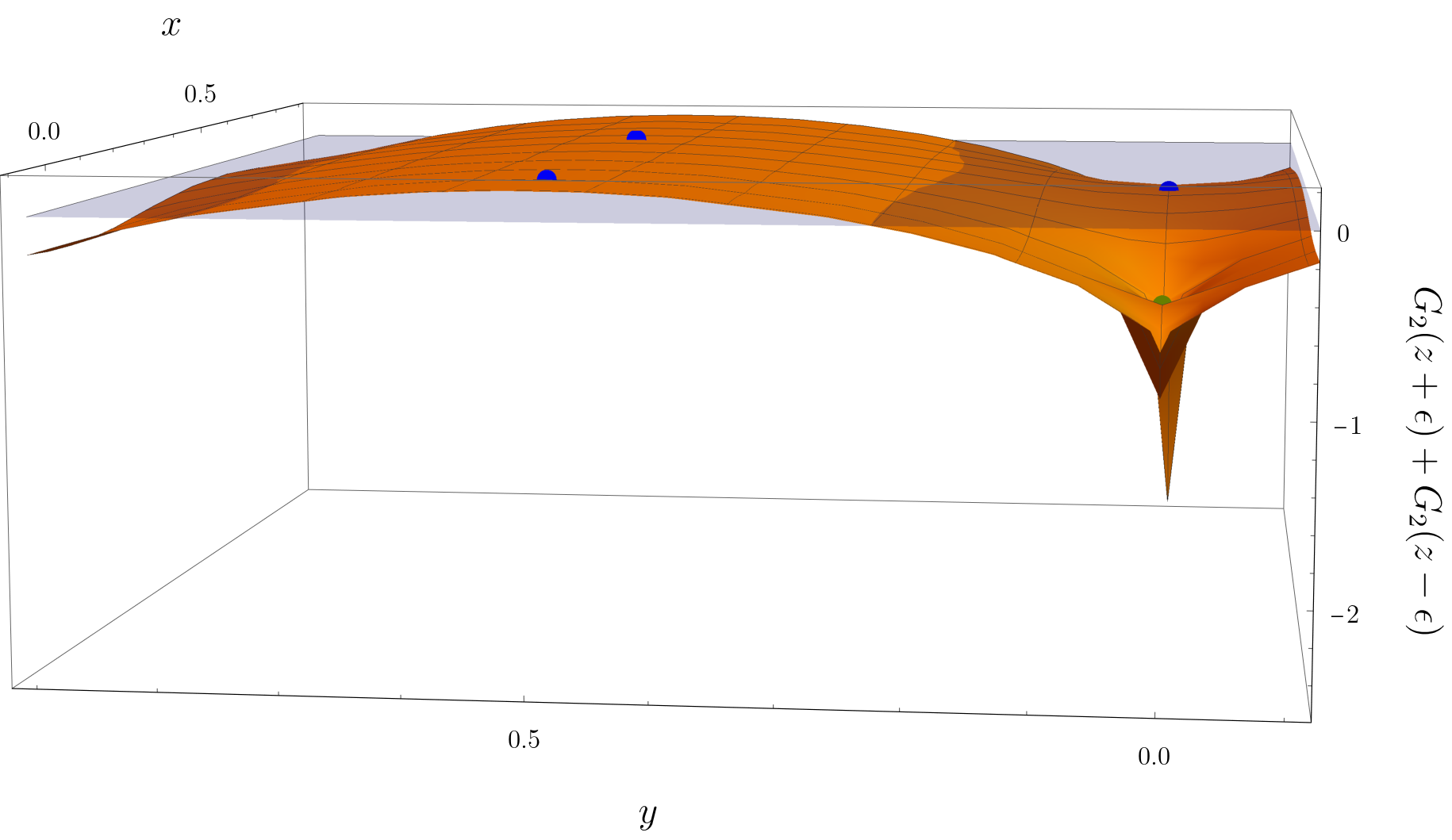}
			\caption{$N=2$}
			\label{fig:t7-e01-N2}
		\end{subfigure}%
		~
		\begin{subfigure}{.5\textwidth}
			\includegraphics[width=\textwidth,keepaspectratio]{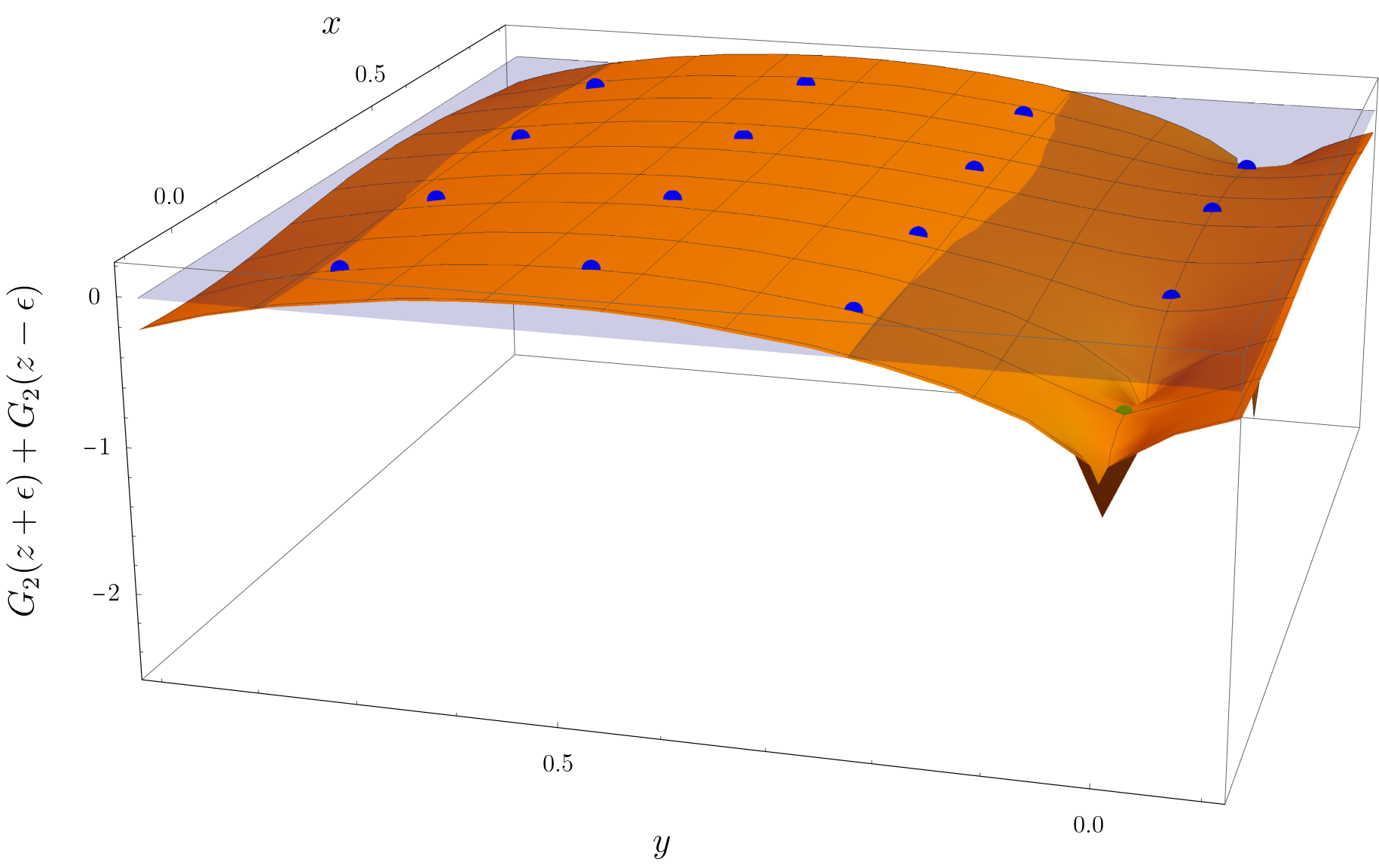}
			\caption{$N=4$}
			\label{fig:t7-e01-N4}
		\end{subfigure}
		\caption{Plot of $G_{\rm tot.}$ for $\tau=2i$ and $\epsilon=0.1$. The blue dots are the ED3 open string landscape points, and the green one $z=0$.}
		\label{fig:t7-e01-N24}
	\end{figure}

\begin{figure}[h!]
	\centering
	\begin{subfigure}{.5\textwidth}
		\includegraphics[width=\textwidth,keepaspectratio]{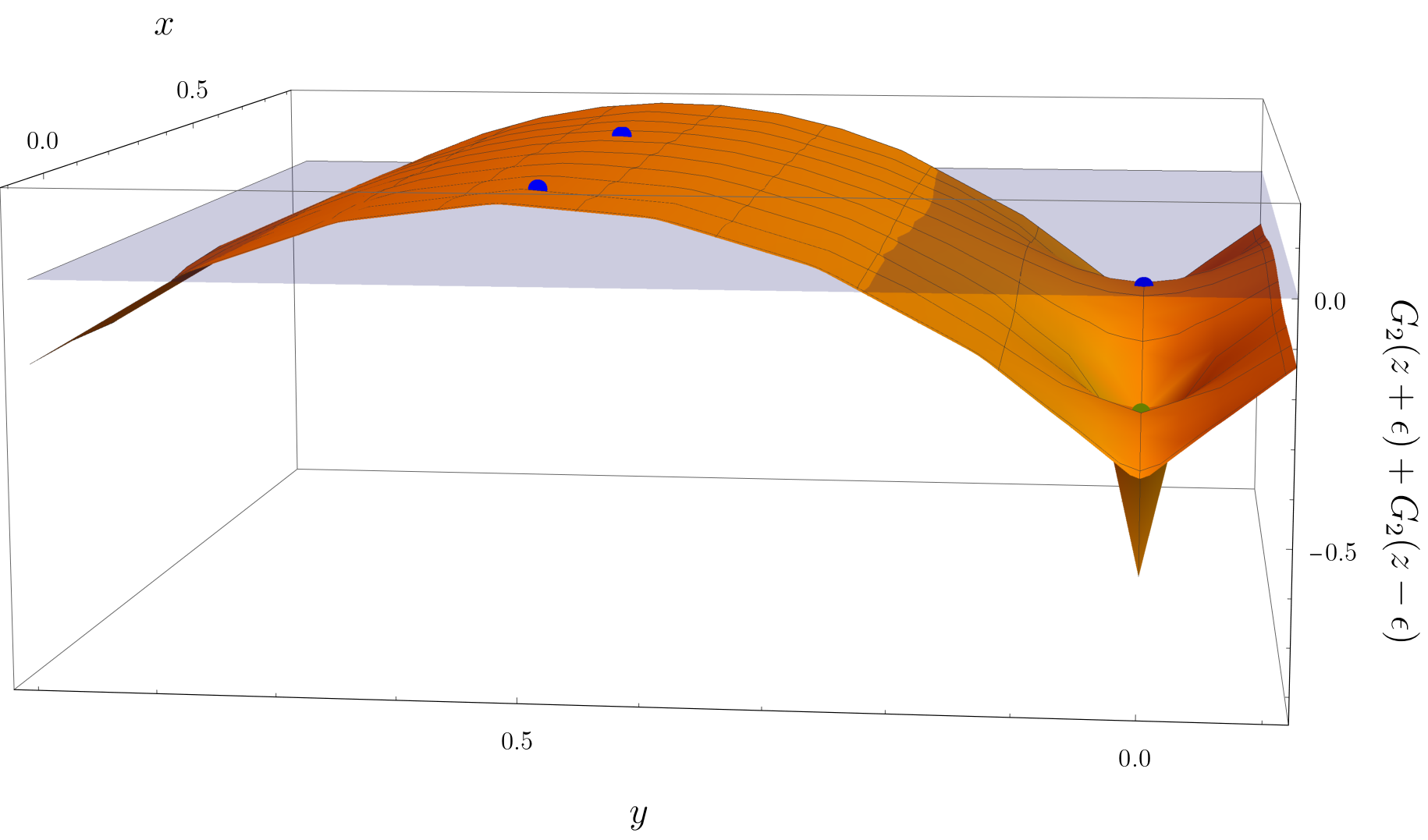}
		\caption{$N=2$}
		\label{fig:t7-e02-N2}
	\end{subfigure}%
	~
	\begin{subfigure}{.5\textwidth}
		\includegraphics[width=\textwidth,keepaspectratio]{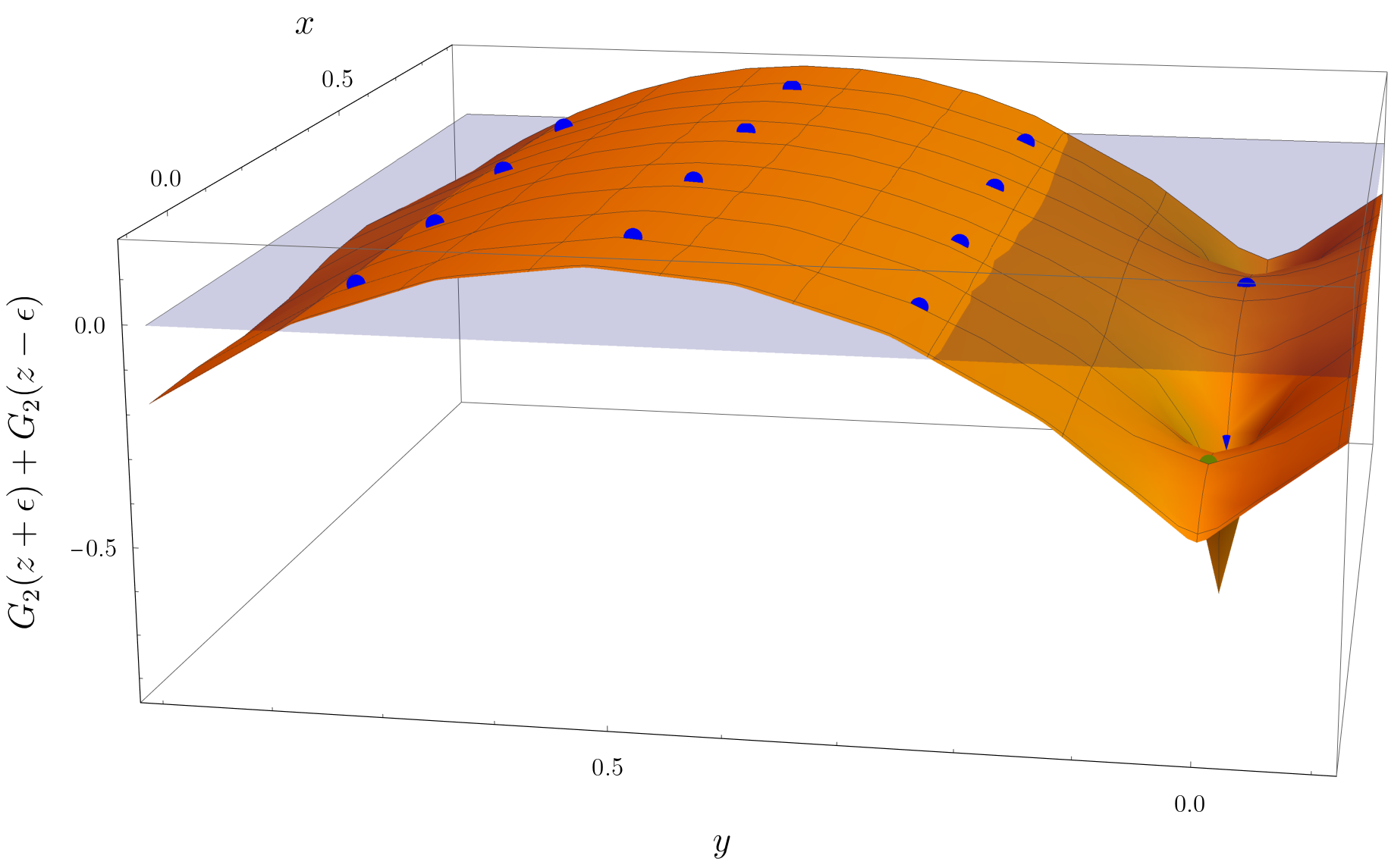}
		\caption{$N=4$}
		\label{fig:t7-e02-N4}
	\end{subfigure}
	\caption{Plot of $G_{\rm tot.}$ for $\tau=2i$ and $\epsilon=0.1$. The blue dots are the ED3 open string landscape points, and the green one $z=0$.}
	\label{fig:t7-e02-N24}
\end{figure}

Since flux quantization in orientifolds of $\IT^6$ requires that $N$ must be at least multiple of 2 (and possibly for 4 or 8, in further orbifolds), the ED3 open string landscape points include at least the four points $z=0,1/2,\tau/2,(1+\tau)/2$. This is non-trivial, since recall the integral of the Green's function integrates to zero over $\IT^2$, hence scans over both positive and negative values. In fact, we have performed extended numerical checks for different values of $\epsilon$ and $\tau$, and have always found that the above condition seems impossible to fulfill. In other words, even if the value of $G_{\rm tot}$ can be made negative at one or even several of these points, there is always at least one of them where $G_{\rm tot}$ is positive. In Figures \ref{fig:t7-e01-N24}, \ref{fig:t7-e02-N24} we provide typical examples for $\tau=2i$, $\epsilon=0.1,\,0.2$, and $N=2,4$.

The pattern is clear and shows that there is no violation of the WGC in this sector of axion charges. Although it could be interesting to have a direct analytical proof of this result, we instead move the discussing the WGC in other closely related charge sectors.

\subsubsection{The fractional ED3/ED$(-1)$ sector}
\label{sec:the-fractional}

The discussion in the previous section makes it clear that the WGC for regular ED3s is satisfied precisely because they are free to move in the $\IT^2$, so that there is always a representative of the charge sector with small enough action. In this section we seek further and consider fractional ED3-branes, which can be stuck at fixed points, and show that the corresponding WGC is violated.

In orientifolds of toroidal orbifolds, in addition to the regular ED3-branes there may be fractional ED3-branes, stuck at the orbifold fixed points. These arise when it is possible to endow the ED3-branes with Chan-Paton indices not in the regular representation of the orbifold group. On general grounds, it is not obvious that one can build models in which D7-branes can be mobile (as we need, to move off the minimum) while admitting fractional ED3-branes stuck at the orbifold points. We postpone this technical discussion, and the construction of an explicit model, to Section \ref{sec:the-orientifold}, and here proceed with those aspects related to the WGC.

For our purposes in this section, it suffices to note that flux quantization in orbifolds ensures that the orbifold points lie at possible ED3 open string landscape positions, so that (suitably magnetized) fractional ED3s maintain their minimal action, without suffering an increase due to the pullbacked B-field. Hence, since these ED3's are stuck at such position, it would naively seem straightforward to find values of  parameters $\tau$, $\epsilon$, such that that $G_{\rm tot.}$ at that particular location is negative. Some obvious examples are given by the Figures above. This would seem to lead to direct violation of the WGC.

However, although the final conclusion is correct, this is not exactly how things work, due to important subtlety. Recall the intuition that fractional branes cannot move off the fixed point because they carry charges under RR fields in the orbifold twisted sector (geometrically, they are secretly wrapped on the cycles collapsed at the orbifold singularity). In other words, they can be regarded as ED3/ED$(-1)$-brane bound states. We are thus dealing with a multi-charge sector, and must hence consider the multi-axion version of the WGC. This is described in terms of the convex hull WGC \cite{Cheung:2014vva,Montero:2015ofa,Rudelius:2015xta}. In the following, we sketch the discussion in the simplified situation that there is only one twisted axion,\footnote{We are also skipping the discussion of the 10d axion, to which ED$(-1)$-brane instantons couple; again they do not significantly change the argument or its conclusion.} since the generalization to several introduces no new features. Also, we abuse language to use the more familiar particle WGC terminology, such as `state' or `mass' (instead of instanton and action).

Consider the situation before the introduction of the 3-form fluxes. We can consider the set of BPS (possibly fractional) ED3/ED$(-1)$-brane bound states, for arbitrary charges. The BPS condition ensures that all these states saturate the WGC condition, and that their charge to mass ratio lie in the extrema region given by the unit ball. Namely, for any rational direction in charge space, there is a BPS state saturating the WGC, namely with unit charge to mass ratio. This is illustrated in Figure \ref{fig:wgca}.

\begin{figure}[htb]
	\centering
	\begin{subfigure}{.49\textwidth}
		\begin{tikzpicture}[scale=2]
		\draw[->,line width=0.8pt,black] (-1.5,0) -- node[above,pos=1,black]{Twisted} (1.5,0);
		\draw[->,line width=0.8pt,black] (0,-1.5) -- node[above,pos=1,black]{Untwisted} (0,1.5);
		\draw[line width=1.5pt,bluscuro] (0,0) circle (1);
		\node at (0,-1.76) {};
		\end{tikzpicture}
		\caption{The 2-axion convex hull WGC for the BPS case. The solid line describes the set of BPS states, saturating the WGC for any rational direction.}
		\label{fig:wgca}
	\end{subfigure}\hfill
   \begin{subfigure}{.49\textwidth}
	\begin{tikzpicture}[scale=2]
	\draw[->,line width=0.8pt,black] (-1.5,0) -- node[above,pos=1,black]{Twisted} (1.5,0);
	\draw[->,line width=0.8pt,black] (0,-1.5) -- node[above,pos=1,black]{Untwisted} (0,1.5);
	\draw[line width=1.5pt,dashed,darkorange] (0,0) circle (1);
	\draw[line width=1.5pt,bluscuro] (0,0) ellipse (0.8 and 1.2);
	\end{tikzpicture}
	\caption{After including backreaction, the curve of former BPS states is deformed away from the unit circle. In the purely untwisted charge direction, the WGC is satisfied, but it is violated in the purely twisted charge direction.}
	\label{fig:wgcb}
  \end{subfigure}
\caption{}
\label{fig:wgc}
\end{figure}

Let us now consider including the backreaction effects. In the direction of the purely untwisted charge, the discussion is as in the previous section, and although the fractional ED3 at the orbifold point does not satisfy the WGC bound, the theory does contain other states satisfying the strict inequality. Hence the curve of former BPS states is deformed outward along that direction, with the deformation controlled by the small parameter $\epsilon$.

In the direction of the purely twisted charge, the charged states are (fractional and regular) ED$(-1)$-branes. Although we have not discussed them in Section \ref{sec:backreactions}, the backreaction effects are straightforward to describe. Since they have no extended dimensions, the ED$(-1)$-branes are insensitive to the warp factor sourced by the induced D3-brane charge; on the other hand, their action is controlled by $g_s^{-1}$, which gets corrected due to the induced D5-brane charge. Promoting its value \eqref{full-local-Z} to the global setup, we find a correction given by \eqref{g-tot}, with a positive definite coefficient $C\sim N|\epsilon|$ from the 5-brane source. In analogy with the argument for ED3s, this correction increases the action for fractional ED$(-1)$ brane instantons, again with the deviation controlled by $\epsilon$. Since we are dealing with twisted charges, there are no representatives in this charge sector in other locations, so there is no charged states satisfying the WGC in this charge sector. Combining results for general rational charge vectors, the WGC diagram looks like Figure \ref{fig:wgcb}. The unit ball is squashed by an amount controlled by $\epsilon$, and there is a violation of the WGC in the direction close enough to the purely twisted axion charge vector.

We can be more precise about Figure~\ref{fig:wgc}. In Figure~\ref{fig:wgca} we are showing all possible states with charges $(n,m)$ under the untwisted and twisted axions given by bound states of ED3-branes wrapped $n$ times on $\IT^2$ and $m$ fractional ED$(-1)$-brane instantons. Since they are BPS they fill out the round circle. In Figure~\ref{fig:wgcb}, the corrections controlled by $\epsilon$ go in opposite directions for the untwisted and twisted directions, and the states fill out the ellipse. Since now the convex hull does not contain the unit ball, there is a violation of the WGC, in particular close to the twisted axion charge vector.
\smallskip

It is worth to stress that such corrections to the action have been computed using the Green function defined in~\eqref{eq:Greenfunc} satisfying~\eqref{the-trick}. Hence the inconsistency with the WGC is clearly related to the fact that the proposed backreaction does not solve the equations of motion, due to the mistreatment of the dynamical tadpole.

As a final remark, note that we have not discussed the values of the Planck scale and axion decay constants. In fact, since we are working in the effective theory of the supersymmetric vacuum, and have simply changed a scalar VEV,  the axion decay constants remain fixed; or more precisely, any  change in the axion decay constant should be encoded in a dependence on the scalars. This would lead to a discussion in terms of the scalar WGC (see \cite{Palti:2017elp}, also \cite{Gonzalo:2019gjp,Gonzalo:2020kke} for variants, and \cite{Palti:2019pca} for the axion version). However, this does not help to satisfy the constraint since e.g. for a single axion the SWGC reads \cite{Palti:2019pca}
\beqa
f^2S^2\,+\,f^2(\partial_\phi S)^2M_P^2\, \leq M_P^2\,.
\eeqa
Hence the  scalar contribution is positive definite and adds to the gravitational contribution. 

The simplest explanation for the non-fulfillment of the WGC in the configuration is thus that it does not provide a consistent background, due to the artificial removal of the dynamical tadpole. In other words, the configuration makes the inconsistency of the background manifest as an incompatibility with quantum gravity, namely with swampland constraint. 

In the next section we build an explicit orientifold with the features described above. Readers not interested in the details are advised to jump to the conclusions in Section \ref{sec:discussion}.

\section{An explicit  $\IZ_2\times \IZ_2$ orientifold example}
\label{sec:the-orientifold}

In this section we build an orientifold of $\IT^6/(\IZ_2\times\IZ_2)$ with mobile D7$_1$-branes which allows for fractional ED3 branes stuck at some fixed points, hence provides an explicit realization of the above ideas. As we will see, the fact that we are interested in D7- and ED3-branes associated to the same 4-cycle in $\IT^6$ makes it non-trivial to allow mobile D7s and stuck ED3s. For the benefit of the reader, we take the opportunity to review the key points of type IIB orientifolds, and some of the main models illustrating them, to better explain our eventual choice of final model realizing mobile D7-branes and stuck ED3-branes. 

\subsection{Choice of discrete torsion}
\label{sec:discrete-torsion}

Orientifolds of $\IT^6/(\IZ_2\times\IZ_2)$ have been studied for decades, and illustrate the wealth of possible discrete parameters in defining orientifolds of orbifold varieties. On one hand, the $\IT^6/(\IZ_2\times\IZ_2)$ orbifold admits a choice of $\IZ_2$ discrete torsion \cite{Vafa:1986wx,Font:1988mk,Vafa:1994rv}. The two resulting models are distinguished by a parameter $\epsilon=\pm 1$ determining the action of the generator $\theta$ of one of the $\IZ_2$'s on the states of the sector twisted by the generator $\omega$ of the second $\IZ_2$ (equivalently, as a relative weight between two disjoint $\text{SL}(2,\IZ)$ orbits of contributions to the torus partition function). Since the twisted sector of $\omega$ contains 2-cycles (collapsed at the orbifold singularities) and 3-cycles (given by the 2-cycles times a 1-cycle in the unrotated $\IT^2$), the choice of discrete torsion determines if the $\theta$ projection keeps the $\omega$-twisted 2-cycles or the 3-cycles. Including a similar analysis for the three different twisted sectors, and the untwisted contributions, the resulting CY threefolds have Hodge numbers $(h_{1,1},h_{2,1})=(3,51)$ for one choice (which we will refer to as without discrete torsion, and denote with $\epsilon=+1$) and $(h_{1,1},h_{2,1})=(51,3)$ for the other choice (which we will refer to as with discrete torsion, and denote with $\epsilon=-1$). We warn the reader that the convention of `with/without' is not uniform in the literature, and that we follow the one used in part of the literature on orientifolds (see later), which is opposite to e.g. \cite{Vafa:1994rv}.

From a geometric perspective, a fractional ED3-brane wrapped on a holomorphic 4-cycle stuck at orbifold fixed points must be secretly wrapped on a collapsed 2-cycle at the singularity. Hence, in order to allow for them, the underlying orbifold model must contain blowup 2-cycles, namely it must be the $(h_{1,1},h_{2,1})=(51,3)$  choice (with discrete torsion or $\epsilon=-1$, in our conventions).

This means that the $\IT^6/(\IZ_2\times\IZ_2)$ orientifold we need is {\em not} the one constructed in \cite{Berkooz:1996km}, which corresponds to the model $(h_{1,1},h_{2,1})=(3,51)$ (without discrete torsion, or $\epsilon=+1$, in our convention). This can be checked by noticing that the Chan-Paton matrices defining the orbifold actions on the D-branes give not a true representation of the orbifold group, but a projective one, as explained in \cite{Douglas:1998xa,Douglas:1999hq} (the `with/without' convention there is opposite to ours, and follows \cite{Vafa:1994rv}). Instead we must focus on models based on the choice $(h_{1,1},h_{2,1})=(51,3)$ (with discrete torsion, or $\epsilon=-1$, in our convention). This may sound troublesome, since models with this choice of discrete torsion tend to have positively charged orientifold planes, and hence require the introduction of antibranes \cite{Angelantonj:1999ms,Klein:2000tf} (see also \cite{Sugimoto:1999tx,Antoniadis:1999xk,Aldazabal:1999jr,Angelantonj:2000xf,Rabadan:2000ma} for examples in other orientifolds). We will later on see how our model does not suffer from this problem.

\subsection{Choice of vector structure}
\label{sec:vector-structure}

When performing the orientifold quotient, there are further discrete choices to consider.\footnote{Actually, the most obvious choice would be the SO/Sp projection. In the discussion below, we assume we take the projection corresponding to negatively charged orientifold planes, to the extent allowed by the other discrete choices (which in some cases force some of these orientifold planes to be positively charged).} In the twisted sectors of a general orbifold element $\IZ_n$, the orientifold usually acts by exchanging oppositely twisted sectors.\footnote{See \cite{Uranga:1999mb} for a notable exception.} This implies that the sector twisted by an order 2 orbifold element $R$ is mapped to itself, and there is a discrete choice of sign for the corresponding orientifold action $\Omega'$ (where the prime indicates that worldsheet parity is in general accompanied by geometric (or other) actions) \cite{Polchinski:1996ry}. As explained in this reference, this manifests in the open string sector of a D$p$-brane as the following condition on the Chan Paton matrices 
\beqa
\gamma_{R,p}\,=\, \pm\, \gamma_{\Omega',p}\,\gamma_{R,p}^{\,T}\, \gamma_{\Omega',p}^{\, -1}\,.
\label{joe-condition}
\eeqa
The choices +/$-$ are (this time, universally) known as with/without vector structure, see \cite{Berkooz:1996iz} (also \cite{Witten:1997bs}) for geometric interpretation underlying the naming. An important aspect is that the choice is correlated with a choice of orientifold action on the closed string sector, hence the sign choice in the open sector must hold for {\em all} D-branes in the model.

A typical choice of Chan-Paton matrices for any of these branes in models without vector structure is, in a suitable basis,
\beqa
\gamma_{R,p}\, =\, \diag( i \id_{n},-i\id_{n})\quad ,\quad \gamma_{\Omega',p}\,=\,\begin{pmatrix} & \id_n \cr \id_n & \end{pmatrix}\;\; {\rm or} \;\; \gamma_{\Omega',p}\,=\,\begin{pmatrix} & i\id_n \cr -i\id_n & \end{pmatrix}\,,
\label{without-vector-structure}
\eeqa
where $n$ denotes the number of D-branes in a given set fixed under the orbifold and orientifold actions, and the two choices of $\gamma_{\Omega',p}$ are symmetric/antisymmetric for the SO/Sp projections. For instance, the 6d $\IT^4/\IZ_2$ model in \cite{Pradisi:1988xd,Gimon:1996rq} corresponds to the above action for $n=16$ on 32 coincident D-branes mapped to themselves by the orbifold and orientifold actions. Similarly,  in the 4d model in \cite{Berkooz:1996km} the orientifold action on each $\IZ_2$ orbifold twist is of this kind (albeit  in not simultaneously diagonalizable way, as befits the projective representation required for the discrete torsion choice of the model). Generalizations to other orbifold groups have also been constructed \cite{Aldazabal:1998mr} (see also \cite{Zwart:1997aj}).

As is clear from the template \eqref{without-vector-structure}, the orientifold acts by exchanging the two different kinds of fractional branes of the underlying $\IZ_2$ orbifold. Hence a consistent orientifold action requires that the orbifold action on D-branes is in the regular representation, namely $\tr \gamma_{R,p}\, =0$, and hence one cannot have stuck fractional D-branes. For our purposes in the main text, this would be fine to allow for mobile D7-branes, but it forbids having stuck ED3-branes. Note that, in the particular case of the model in \cite{Berkooz:1996km}, this also agrees with our earlier discussion of the absence of fractional 4-cycles for that choice of discrete torsion.

Hence, for our purposes in the main text, we are interested in models where the orientifold action on an orbifold element rotating the first complex plane is with vector structure. Models with vector structure have been considered, starting from \cite{Polchinski:1996ry} (see \cite{Klein:2000hf} for 4d examples) and they involve an extra subtlety. In the D9-brane description, the orbifold fixed points also fixed under this orientifold action with vector structure, have positive RR charge. Hence, as shown in \cite{Polchinski:1996ry}, a consistent supersymmetric model can be achieved only if 8 of the 16 fixed points have orientifold action with vector structure, and the other 8 have orientifold action without vector structure (with the difference implemented by a suitable Wilson line). The model thus contains D9-branes, but no D5-branes. The model has a T-dual with D5-branes and no D9-branes, which had been constructed in \cite{Dabholkar:1996zi}. We now turn to the construction of this 6d model, but in terms of D7-branes, to later employ it to build a 4d model with mobile D7-branes and admitting stuck ED3s.

The construction of the 6d model is as follows. Take $\IT^4$ parameterized by $(z_1,z_2)$ with $z_i=x_i+\tau_i y_i$, and $x_i, y_i$ with periodicities 1. We have a $\IZ_2$ orbifold action generated by $\theta:(z_1,z_2)\to (-z_1,-z_2)$, with 16 orbifold fixed points at the locations $x_i=0, \frac 12$, $y_i=0, \frac 12 $. We orientifold by  $\Omega R_1 (-1)^{F_L}$ with $R_1:(z_1,z_2)\to\left(-z_1+\frac 12,z_2\right)$. This leads to 4 O7$_1$-planes (which we take negatively charged), located at $x_1=\frac 14,\frac 34$, $y_1=0, \frac 12$, and spanning $z_2$. There are 32 D7$_1$-branes, whose distribution in the $z_1$ plane should respect the symmetries, and will be discussed later on. The orbifold fixed points do not sit on top of the O7$_1$-planes, so the orientifold action exchanges them. The orientifold group also includes the element $\Omega R_1 \theta (-1)^{F_L}$, which is however freely acting since $R_1\theta:(z_1,z_2)\to \left(z_1+\frac 12,-z_2\right)$. Hence, there are no crosscap RR tadpoles, namely no O7$_2$-planes, and hence no D7$_2$-branes need to be introduced. This reproduces the cancellation of untwisted tadpoles, with only one kind of brane, as in the T-duals \cite{Dabholkar:1996zi,Polchinski:1996ry}. On the other hand, since the orientifold planes do not coincide with the orbifold fixed points, they do not induce twisted RR tadpoles. Hence the D7$_1$-branes can be located anywhere in the $z_1$-plane, for instance as 8 independent D7-branes with their corresponding orbifold and orientifold images. If they are located on top of an orbifold or orientifold fixed point, their symmetry is enhanced. For instance, locating 16 D7-branes on top of an orientifold plane and 16 on top of its orbifold image, one gets  $\SO(16)$ vector multiplets of 6d $\NN=1$, with one adjoint hypermultiplet. If we locate 16 D7-branes on top of an orbifold point, and 16 at its orientifold image, cancellation of RR disk twisted tadpoles enforces
\beqa
\tr \gamma_{\theta,7_1}\, =\, 0\; \rightarrow \; \gamma_{\theta,7_1}\, =\, \diag(\id_8,-\id_8)\,,
\label{tadpole-with-vs}
\eeqa
at each of the two orbifold fixed points. In this case the gauge group is $\U(8)^2$, with two hypermultiplets in the $(\fund,\antifund)$. Although sitting at the orbifold fixed point, the D7$_1$-branes can be move off into the bulk, and this corresponds to Higgsing with the bifundamental down to less symmetric patterns, possibly down to the generic $\U(1)^8$. As a related comment, note that the equality of +1 and -1 entries in \eqref{tadpole-with-vs} is not enforced by the orientifold action (which is merely mapping one orbifold fixed point to the other), but by the disk RR tadpole condition. This implies that it is perfectly consistent to have a D3-brane wrapped on the directions $z_2$ and sitting at an orbifold fixed point in $z_1$ (with another D3-brane at its orientifold image position) with
\beqa
\tr\gamma_{\theta,3}\neq 0 \; \rightarrow\,  {\rm e.g. }\;\gamma_{\theta,3}\, =\, 1\,.
\label{the-stuck}
\eeqa
This D3-brane sources a twisted tadpole, but there are non-compact dimensions transverse to it in which the flux lines can escape to infinity. The fact that the D3-brane sources this charge implies it cannot be moved off the orbifold fixed point. Indeed, the open string sector does not contain any matter fields for the choice \eqref{the-stuck}. This wrapped D3-brane corresponds to a BPS string in the 6d theory, and is to become a stuck ED3 in the upcoming 4d model, by wrapping its two dimensions on the extra $\IT^2$.

\subsection{The 4d model}
\label{sec:the-model}

It is now easy to combine different above ingredients to build a 4d model with mobile D7-branes and admitting stuck ED3s. We consider a factorized $\IT^6$ parameterized by $(z_1,z_2,z_3)$ with $z_i=x_i+\tau_i y_i$, and $x_i, y_i$ with periodicities 1. We mod out by the $\IZ_2\times\IZ_2$ orbifold action generated by $\theta:(z_1,z_2,z_3)\to (-z_1,-z_2,z_3)$ and $\omega:(z_1,z_2,z_3)\to (z_1,-z_2,-z_3)$. This leads to the familiar 16$\times$3 orbifold fixed planes, and as explained in Section \ref{sec:discrete-torsion}, we choose the model leading to $(h_{1,1},h_{2,1})=(51,3)$ (i.e. with discrete torsion or $\epsilon=-1$, in our convention).

We now perform an orientifold $\Omega R_1(-1)^{F_L}$, with $R_1:(z_1,z_2,z_3)\to \left(-z_1+\frac 12,z_2,z_3\right)$. As explained in Section \ref{sec:vector-structure}, this leads to 4 O7$_1$-planes located at $x_1=\frac 14,\frac 34$, $y_1=0, \frac 12$, and spanning $z_2$. On the other hand, $\Omega R_1\theta(-1)^{F_L}$ and $\Omega R_1\theta\omega(-1)^{F_L}$ act with a shift in the coordinate $z_1$, hence are freely acting and do not introduce O7$_2$- and O7$_3$-planes. Finally $\Omega R_1\omega(-1)^{F_L}$ acts geometrically as $R_1\omega:(z_1,z_2,z_3)\to \left(-z_1+\frac 12,-z_2,-z_3\right)$, and leads to 64 O3-planes, at $x_1=\frac 14,\frac 34$, $y_1=0, \frac 12$, $x_i,y_i=0, \frac 12$ for $i=2,3$. Note that the O7$_1$ planes exchange the $\theta$-fixed orbifold points (and similarly for the $\omega\theta$-fixed points), but maps each $\omega$-fixed plane to itself (and similarly for the O3-planes). In particular, notice that there are points which are simultaneously fixed under the $\omega$ action and the O3-plane (or O7$_1$-plane) action.

We must now specify the discrete choices for these orientifold actions, to achieve the desired result. We take negatively charged O7$_1$-planes, so as to have a total of 32 D7$_1$-branes, as counted in the covering space. Since we seek to have mobile D7$_1$-branes and stuck ED3-branes in $z_1$, we need the action of $\Omega R_1(-1)^{F_L}$ on the $\theta$ orbifold to be with vector structure, just as in the last 6d example discussed above. On the other hand, the action of $\Omega R_1(-1)^{F_L}$ on the $\omega$ orbifold cannot be with vector structure, since this would lead to positively charged O3-planes, whose RR charge cannot be canceled in a supersymmetric way. Hence this sector should have Chan Paton matrices without vector structure. The orientifold action on the $\theta\omega$ sector follows from the above, and is without vector structure. 

Notice that this pattern matches the observation in \cite{Angelantonj:1999ms} that in orientifolds of $\IT^6/(\IZ_2\times \IZ_2)$ there are three discrete sign choices $\epsilon_i$ determining the orientifold action on the corresponding orbifold element, morally $\epsilon_i=+1$ (resp. $\epsilon_i=-1$) implies the corresponding orientifold planes are negatively (resp. positively) charged.
These signs are correlated with the discrete torsion parameter $\epsilon$ by $\epsilon_1\epsilon_2\epsilon_3=\epsilon$. Our model has discrete torsion $\epsilon=-1$, and hence requires that at least one orientifold action has $\epsilon_i=-1$. However, the model cleverly evades the need to introduce positively charged orientifolds planes, because the $\epsilon_i=-1$ action corresponds to the $\theta$ orbifold sector, where the orientifold action is freely acting and no actual orientifold planes appear. The two $\epsilon_i=+1$ sectors are the $\theta\omega$ sector, without orientifold planes, and the $\omega$, with negatively charged O3-planes.

To make the above description more explicit, let us describe the Chan-Paton action on the 32 D3-branes. In the actual model in the main text, the D3-branes will actually be replaced by 3-form fluxes, see Section \ref{sec:the-fluxes}, so they are here used just to illustrate the effect of discrete choices in open string sectors. 

We consider 16 D3-branes located at and O3-plane, and 16 at the image under the orbifold action $\theta$ (or $\theta\omega$). Each set is mapped to itself by the action $\omega$ and by $\Omega'\equiv \Omega R_1(-1)^{F_L}$, which should be represented by matrices $\gamma_{\omega,3}$, $\gamma_{\Omega',3}$ of the form \eqref{without-vector-structure}. On the other hand, if we include the 16+16 set in a single matrix $\gamma_{\theta,3}$ to describe the action of $\theta$, its interplay with $\Omega R_1(-1)^{F_L}$ should be with vector structure. Finally, recall that the matrices $\gamma_{\theta,3}$, $\gamma_{\omega,3}$ should provide a non-projective representation of the orbifold group, due to the choice of discrete torsion. A simple choice satisfying these properties is
\beqa
\gamma_{\theta,3}\, =\, \begin{pmatrix} & \id_{16} \cr \id_{16} & \end{pmatrix} \; ,\; \gamma_{\omega,3}\,=\, \diag(i\id_8,-i\id_8;i\id_8,-i\id_8) \; ,\;
\gamma_{\Omega',3}\,=\, \begin{pmatrix}  & i\id_8 & &  \cr  -i\id_8 & & & \cr & & &  i\id_8 \cr  & & -i\id_8 &\end{pmatrix} \nonumber\,.
\eeqa
Although $\gamma_{\theta,3}$ and $\gamma_{\omega,3}$ commute, we have not diagonalized the former so as to maintain the 16+16 split manifest. The matrices satisfy \eqref{joe-condition} with $+$ sign for $\gamma_{\theta,3}$ and $-$ sign for $\gamma_{\omega,3}$. We have chosen antisymmetric $\gamma_{\Omega',3}$ corresponding to an Sp projection. 

For D7$_1$ branes, we choose to locate 8 on top of each O7$_1$-plane, which we recall are exchanged pairwise by $\theta$. The Chan-Paton matrices in the corresponding 8+8 set is similar to the above, but with symmetric $\gamma_{\Omega',7}$, namely
\beqa
\gamma_{\theta,7_1}\, =\,  \begin{pmatrix} & \id_{8} \cr \id_{8} & \end{pmatrix} \quad ,\quad \gamma_{\omega,7_1}\,=\, \diag(i\id_4,-i\id_4;i\id_4,-i\id_4)\quad, \quad\gamma_{\Omega',7_1}&=& \begin{pmatrix}  & \id_4 & &  \cr  \id_4 & & & \cr & & &  \id_4 \cr  & & \id_4 \nonumber &\end{pmatrix} \,.
\eeqa

\subsection{Introducing 3-form fluxes}
\label{sec:the-fluxes}

As already mentioned, and is clear in the main text, the model must include NSNS and RR 3-form fluxes, which contribute to the RR 4-form tadpole cancellation. In particular, in the normalization \eqref{flux-def}, we have
\beqa
N_{\rm flux}\,=\, \frac{1}{(2\pi)^4\alpha'{}^2}\int_{\IX_6} F_3\wedge H_3\, =\, 2N^2\,.
\eeqa
The RR tadpole condition is
\beqa
N_{\rm flux}\, +\,N_{\rm D3}\, =\, 32\,,
\eeqa
where $N_{\rm D3}$ is the number of D3-branes, as counted in the covering space. Hence, the condition for the flux contribution not to overshoot\footnote{Notice that a moderate overshoot can actually be allowed, and still maintain supersymmetry, if one includes suitably magnetized D9-branes, as implemented in \cite{Marchesano:2004yq,Marchesano:2004xz} to solve a similar overshooting problem in \cite{Blumenhagen:2003vr,Cascales:2003zp,Cascales:2003pt}.} the RR tadpole is that $N\leq 4$.

One may fear that this bound is too small for an orientifold of $\IT^6/(\IZ_2\times\IZ_2)$, due to flux quantization conditions. As mentioned in the main text, it was argued in \cite{Frey:2002hf} that in orientifolds of $\IT^6$, NSNS and RR fluxes must be quantized in multiples of 2, if the model has all negatively charged O3-planes, while odd quanta are allowed only if other (positively charged) exotic O3-planes are included. In addition, the $\IZ_2$ orbifold projections in general allow for smaller 3-cycles than in the underlying $\IT^6$, leading to more strict quantization conditions. In fact, as shown in \cite{Blumenhagen:2003vr,Cascales:2003zp} the orientifold of the $\IT^6/(\IZ_2\times\IZ_2)$ with $(h_{11},h_{21})=(3,51)$ (without discrete torsion, or $\epsilon=1$, in our convention) requires 3-form fluxes to be quantized in multiples of 8, leading to an overshoot of their RR tadpole contribution. However, happily, for the $\IT^6/(\IZ_2\times\IZ_2)$ we are actually using, with $(h_{11},h_{21})=(51,3)$  (with discrete torsion, or $\epsilon=-1$, in our convention), it was shown in \cite{Blumenhagen:2003vr} that 3-form fluxes must be quantized in multiples of 4. Thus, the minimum amount of flux available for this model leads to $N=4$, which precisely saturates the RR tadpole cancellation, without need of D3-branes. 

We would like to finish with an important observation, which relates flux quantization with the `energetics' of the stuck ED3-branes of interest in the main text. The 3-form fluxes \eqref{flux-def} are clearly invariant under the orbifold and orientifold transformations.
However, this invariance is not manifest once we write down explicit expressions for the NSNS 2-form gauge potential. For instance, let us redefine the origin in the $z^1$-plane, so that the origin $z^1=0$ corresponds to an orientifold plane, as in the main text. Then, we may write
\beqa
B&= &4\pi^2\alpha'\, N\, \left(\, x^1\, dx^2\wedge dx^3\, +\, y^1\, dy^2\wedge dx^3\,\right)\,,
\label{local-b}
\eeqa
c.f. \eqref{bfield-ond71}. This is invariant under the orientifold action, but not invariant under the orbifold action (which in the $z_1$-plane acts as a reflection with respect to e.g. $(x^1,x^2)=\left(\frac 14,0\right)$. Clearly this is just because  \eqref{local-b} holds in a local patch, and we are allowed to make gauge transformations among different patches. Hence, near $(x^1,x^2)=\left(\frac 14,0\right)$ we may fix a different gauge and represent the same $H_3$ with
\beqa
B &= &4\pi^2\alpha'\, N\, \left(\, \left(x^1-\textstyle{\frac {1}{4}}\right)\, dx^2 \wedge dx^3\, +\, y^1\, dy^2\wedge dx^3\,\right)\,,
\label{local-b-prime}
\eeqa
which is invariant under the orbifold, but not the orientifold action. The question now is, if we consider ED3-brane instantons stuck at the orbifold fixed point, which of the two expressions for the B-field should we consider? This is relevant, because its pullback on the ED3 worldvolume provides a contribution to the ED3 action, and hence seems to have an impact on whether or not the WGC is satisfied. The answer is simply that both expressions are valid, if we consider not just a given ED3, but rather the whole set of magnetized ED3s, with different magnetization quanta. Indeed, the shift in the B-field upon the gauge transformation can be translated in a change in the magnetization of an ED3 by an amount
\beqa
-\frac 14\,N\, dx^2\wedge dx^3\,.
\eeqa
 This can be absorbed by a properly quantized worldvolume magnetic flux precisely thanks to the 3-form flux quantization condition $N=4$ (or a multiple thereof, in general). 
 
 An equivalent description is in terms of axion monodromy, with the `axion' given by position of the ED3-branes; considering the full tower of magnetized ED3-branes, the tower at $z^1=\frac 14$ is identical to the  tower at $z^1=0$ modulo ED3-branes with different magnetization $F_2=n\, dx^2\wedge dx^3$ changing as $n\to n-1$. Another equivalent description is in the language of \cite{Gomis:2005wc} c.f. Section \ref{sec:the-landscape}, as follow. In terms of the B-field \eqref{local-b}, there is an open string landscape of BPS ED3's at points $x^1N\in\IZ$, $y^1N\in\IZ$; hence, for $N=4$, the orbifold fixed point $z^1=\frac 14$ is one of the open string landscape points where some ED3 with suitable magnetization can cancel the corresponding B-field. This cancellation is made manifest in the alternative local expression \eqref{local-b-prime}. Notice that a similar mechanism is exploited for the D7$_1$-branes so that their distribution in sets of 8 on top of the O7$_1$-planes, as discussed  in Section \ref{sec:the-model}, remains a valid supersymmetric background in the presence of fluxes.
 
\section{Discussion and Conclusions}
\label{sec:discussion}

In this paper we have considered the backreaction of supersymmetry breaking effects, and the corresponding dynamical tadpole, in explicit examples of type IIB toroidal orientifold. We have shown that the resulting configurations seem to violate the WGC for certain axions. We have argued that the underlying problem is due to the unphysical assumption of ignoring the effects of the dynamical tadpoles on the 4d spacetime configuration, restricting the backreaction to the internal space. Hence, these are examples of theories in which dynamical tadpoles manifest as direct incompatibility with quantum gravity, via swampland constraints.

These examples and the above interpretation open up many new avenues, among others:
\begin{itemize}
	\item Our source of supersymmetry breaking is based on moving slightly off the minimum of an otherwise supersymmetric theory. It would be nice to carry out the arguments in this paper in a genuinely non-supersymmetric model.
	\item It would be interesting to find models where the spacetime dependence sourced by the dynamical tadpole can be solved, and to address the formulation of the WGC in those backgrounds. In particular, it may well be possible that the WGC does not hold in its usual formulation. For instance, the usual black hole arguments for the WGC for particles is based on the stability of remnants, a feature which is sensitive to new effects if one is considering e.g. time-dependent configurations.
	\item We have also encountered models where the dynamical tadpole does not seem to lead to violation of the WGC. It would be interesting to explore if they violate some other swampland constraint. Conversely, these models could potentially be used to uncover new swampland constraints not considered hitherto.
	\item Cancellation of topological tadpoles (such as RR tadpoles), which are often associated to cancellation of anomalies in the spacetime theory, or on suitable probes \cite{Uranga:2000xp}. The ED3 and ED$(-1)$-brane instantons in our examples are reminiscent of probes of the dynamical tadpoles, albeit in a dynamical rather than topological way. It would be interesting to explore the interplay of dynamical tadpoles and probes in more general setups.
\end{itemize}
These and other related questions seem capable of shedding new light in the long-standing problem of dynamical tadpoles in string theory.
We hope to come back to them in future work.

\section*{Acknowledgments}
We are pleased to thank A. Herráez, L. Ib\'anez, F. Marchesano, M. Montero and I. Valenzuela for useful discussions. This work is supported by the Spanish Research Agency (Agencia Estatal de Investigaci\'on) through the grant IFT Centro de Excelencia Severo Ochoa SEV-2016- 0597, and by the grant PGC2018-095976-B-C21 from MCIU/AEI/FEDER, UE. The work by A.M. is supported by ``la Caixa'' Foundation (ID 100010434) with fellowship code LCF/BQ/IN18/11660045 and from the European Unions Horizon 2020 research and innovation programme under the Marie Sklodowska-Curie grant agreement No. 713673.

\newpage

\appendix

\section{$Z$-minimization and $a$-maximization from WGC}
\label{app:z-minimization}

In the main text we have shown that in certain string models, configurations away from the vacuum lead to an uncanceled dynamical tadpole which manifests as a non-fulfillment of the WGC. Hence, the condition to satisfy the WGC by minimizing the action of suitable charged states turns out to be equivalent to minimization of the scalar potential. In this appendix we explain how this idea explains the condition of Z-minimization in the context of AdS vacua in holography. We point out that in this context the deviation from the vacuum is not a modulus or light scalar direction, but rather involves modes with masses comparable with the cutoff, i.e. the KK scale; this suggests a more general validity of our arguments beyond their use in effective field theory.

\subsection{Overview of $Z$-minimization}

In this section we recall the key ideas in \cite{Martelli:2005tp,Martelli:2006yb}. Consider  the AdS/CFT duality between 4d $\NN=1$ quiver gauge theories, obtained from D3-branes at  a toric CY threefold singularity $\IY_6$, and type IIB string theory on AdS$_5\times \IX_5$, with the horizon $\IX_5$ given by the base of the real cone $\IY_6$. The CY condition of $\IY_6$ implies that $\IX_5$ is Sasaki-Einstein, and has
at least one $\U(1)$ symmetry (dual to the SCFT $\U(1)_R$-symmetry). It is generated by the Reeb vector, obtained from the complex structure $J$ of $\IY_6$ by
\begin{equation}
\xi=J\left(r \frac{\de}{\de r}\right)\coma
\end{equation}

Using the condition that $\IX_5$ is Einstein, it is possible to fix the normalization of the Ricci tensor as
\begin{equation}
\mathcal{R}_{mn}=4g_{mn}\fstop
\end{equation} 
This Ricci tensor can be obtained extremizing the Einstein-Hilbert action, which can be recast as the volume of $\IX_5$ \cite{Martelli:2005tp}
\begin{equation}
S[g]=\int_{\IX_5}d^5x\sqrt{g}\left(\mathcal{R}_{\IX_5}-12\right)\,=\,8\text{Vol}(\IX_5)
\end{equation}
which in turn depends only on the Reeb vector $\xi$. This means that the problem of finding the metric for the Sasaki-Einstein manifold reduces to the minimization of the volume with respect to the Reeb vector. Although this is the starting point for the WGC discussion in the next section, for completeness we close this flash review with the expression of this volume in terms of the toric data.

For toric $\IY_6$, the Sasaki-Einstein manifold $\IX_5$ has at least a $\U(1)^3$ isometry. Let us introduce the 3d vectors defining the toric fan data of $\IY_6$; since they lie on a plane, they are of the form $v_i=(1,w_i)$, with $w_i$ giving the toric diagram of the geometry. The computation of the volume of $\IX_5$ is done using the Duistermaat-Heckman formula and via localization, which boils down to simple closed formulas in the toric case. We write down the coordinates of $\xi=(3,b_2,b_3)$ as
\begin{equation}
\text{Vol}\left(\IX_5\right)\equiv\frac{\pi^3}{3}\sum_{i=1}^d \frac{\det\left(v_{i-1},v_i,v_{i+1}\right)}{\det\left(\xi,v_{i-1},v_i\right)\det\left(\xi,v_i,v_{i+1}\right)}\coma
\label{eq:volX51}
\end{equation}
The quantity $Z$ is defined as the volume of a Sasaki-Einstein manifold, relative to that of the round sphere. It is an algebraic number given by
\begin{equation}
Z(b_2,b_3)\equiv\frac{1}{(2\pi)^3}\text{Vol}\left(\IX_5\right)=\frac{1}{24}\sum_{i=1}^d \frac{\det\left(v_{i-1},v_i,v_{i+1}\right)}{\det\left(\xi,v_{i-1},v_i\right)\det\left(\xi,v_i,v_{i+1}\right)}\coma
\label{eq:Zfun}
\end{equation}
The AdS solution thus corresponds to the configuration which minimizes this quantity with respect to the Reeb vector, a procedure known as $Z$-minimization. Incidentally, this provides the gravity dual of the $a$-maximization \cite{Intriligator:2003jj} in the holographic 4d $\NN=1$ SCFT.

In the following section we will look at particles obtained by D3-branes wrapped on a $3$-cycle. Such cycle $\Sigma_i$ has a volume that also can be expressed as a function of the Reeb vector and, in the toric case, using the fan data of $\IY_6$.  It is another result in~\cite{Martelli:2005tp} that the volume of a $3$-cycle as a function of $\xi$ is
\begin{equation}
\text{Vol}\left(\Sigma_i\right)\equiv 2\pi^2 \frac{\det\left(v_{i-1},v_i,v_{i+1}\right)}{\det\left(\xi,v_{i-1},v_i\right)\det\left(\xi,v_i,v_{i+1}\right)}\fstop
\label{eq:volS1}
\end{equation}
From this expression it is possible to show that\footnote{For details of the general proof for toric Sasaki-Einstein manifold, the reader can have a look at~\cite{Martelli:2005tp}.} 
\begin{equation}
\text{Vol}\left(\IX_5\right)=\frac{\pi}{6}\sum_{i=1}^d \text{Vol}\left(\Sigma_i\right)\fstop
\label{eq:relvolX5S1}
\end{equation}
Finally, before proceeding with the WGC, it is better to define $R$ the radius of the AdS space and of the internal manifold $\IX_5$. Eq.~\eqref{eq:volX51} and~\eqref{eq:volS1} become
\begin{equation}
\begin{split}
\text{Vol}\left(\IX_5\right)&=\frac{\pi^3R^5}{3}\sum_{i=1}^d \frac{\det\left(v_{i-1},v_i,v_{i+1}\right)}{\det\left(\xi,v_{i-1},v_i\right)\det\left(\xi,v_i,v_{i+1}\right)}\coma\\
\text{Vol}\left(\Sigma_i\right)&= 2\pi^2R^3 \frac{\det\left(v_{i-1},v_i,v_{i+1}\right)}{\det\left(\xi,v_{i-1},v_i\right)\det\left(\xi,v_i,v_{i+1}\right)}\fstop
\end{split}
\end{equation}
This means that Eq.~\eqref{eq:relvolX5S1} is
\begin{equation}
\text{Vol}\left(\IX_5\right)=\frac{\pi R^2}{6}\sum_{i=1}^d \text{Vol}\left(\Sigma_i\right)\fstop
\end{equation}
Notice, moreover, that, we can define the Z-function normalizing the volume with respect to a 5-sphere of radius $R$ and still get the same expression as in Eq.~\eqref{eq:Zfun}. 

\subsection{$Z$-minimization from WGC}

We now show that the above condition of $Z$-minimization is equivalent to the requirement that the WGC is satisfied for a suitable set of charged states in the AdS theory. For this purpose, we consider type IIB theory compactified on AdS$_5\times \IX_5$, with the volume of $\IX_5$ relative to that of $\IS^5$ (of same radius as the AdS, $R$) given by the function $Z(\xi)$, which is minimized at the vacuum.

We now consider a set of states whose masses depend on the Reeb vector. As explained above, we can take D3-branes wrapped on 3-cycles $\Sigma_i$ of $\IX_5$. The ratio of the masses $m_i/m_{i;0}$ of such state for a general trial Reeb vector, and for the vacuum one is
\begin{equation}
\frac{m_i}{m_{i;\,0}}=\frac{\text{Vol}(\Sigma_i)}{\text{Vol}_{\text{min}}(\Sigma_i)}\fstop
\label{mass-ratio}
\end{equation}
In order to express it in a WGC format, we need to obtain the gauge couplings of the $\U(1)_R$ under which these are charged. By dimensional reduction from 10d we have
\begin{equation}
g^{-2}\,=\,M_s^8\, g_s^{-2}\,\text{Vol}(\IX_5)\,R^2\coma
\end{equation}
where the last $R^2$ is just a standard normalization factor.

We can also compute the 5d Planck mass by reduction from 10d to get
\beqa
M_{P,\,5}^3\,=\,M_s^8\, g_s^{-2}\,\text{Vol}(\IX_5)\,.
\eeqa
They are related by 
\beqa
g\, M_{P,\,5}^{\frac 32}\,=\, R^{-1} \,,
\eeqa
independently of $\text{Vol}(\IX_5)$. This is useful, since in this system the free parameters in the Reeb vector are not actual moduli in this configuration (their masses are of the order of the KK scale), so their change is not really a deformation in a given effective theory. Hence, it is questionable to use the same or different values for $g$ and $M_{P,\,5}$ in comparing the vacuum and configurations away from it. The above relation allows us to circumvent this discussion and proceed to the result.

We now use that the wrapped D3-branes are BPS states at the vacuum and satisfy
\beqa
m_{i;\,0}\,=\,g\, Q\,M_{P,\,5}^{\frac 32}\,=\,\frac{Q}{R}\fstop
\eeqa
where $Q$ denotes its charge under $\U(1)_R$. Hence at the configuration away from the vacuum we have
\beqa
m_i\,=\,g\,Q\,M_{P,\,5}^{\frac 32}\,\frac{\text{Vol}(\Sigma_i)}{\text{Vol}_{\text{min}}(\Sigma_i)}\fstop
\label{eq:mm0bargg0WGC2}
\eeqa
Due to the convexity of the volume functions with respect to the Reeb vector $\xi$~\cite{Martelli:2005tp,Martelli:2006yb} (see also~\cite{Butti:2005vn,Butti:2006nk,Eager:2010yu}), the only way to satisfy the WGC is to take the value $\xi=\xi_{\text{min}}$. In other words, we recover the Z-minimization condition from WGC considerations.

\newpage
\bibliographystyle{JHEP}
\bibliography{mybib}

\end{document}